\pdfoutput=1

\documentclass[review,3p,times]{elsarticle}
\usepackage{tikz}
\usepackage{bm}
\usepackage{amsmath}
\usepackage{mathtools}
\usepackage{dirtytalk}
\usepackage{derivative}
\usepackage{threeparttable}
\usepackage{lscape}
\usepackage{float}
\usepackage[pagewise]{lineno}
\usepackage{booktabs}
\usepackage{threeparttable}
\usepackage{caption}
\usepackage{pgfplots}
\usepackage{subcaption}
\usepackage[short,nocomma]{optidef}
\usepackage{hyperref}

\pgfplotsset{compat=1.18}
\newcommand{\tran}{{\mkern-1.5mu\mathsf{T}}}
\biboptions{sort&compress}

\begin{document}

\begin{frontmatter}

\author[1]{Sebastián Espinel-Ríos\fnref{fn1}}

\author[1]{Gerrich Behrendt\fnref{fn1}}

\author[1]{Jasmin Bauer}

\author[2]{Bruno Morabito}

\author[3]{Johannes Pohlodek}

\author[1]{Andrea Schütze}

\author[3]{Rolf Findeisen}

\author[1]{Katja Bettenbrock}

\author[1]{Steffen Klamt\corref{cor1}}
\ead{klamt@mpi-magdeburg.mpg.de}

\cortext[cor1]{Corresponding author}
\fntext[fn1]{These authors contributed equally. SER contributed to the modeling and optimization methods. GB contributed to the genetic engineering and experimental methods.}

\affiliation[1]{organization={Analysis and Redesign of Biological Networks, Max Planck Institute for Dynamics of Complex Technical Systems},
            addressline={Sandtorstraße 1}, 
            city={Magdeburg},
            postcode={39106}, 
            state={Sachsen-Anhalt},
            country={Germany}}

\affiliation[2]{organization={Yokogawa Insilico Biotechnology GmbH},
            addressline={Meitnerstraße 9}, 
            city={Stuttgart},
            postcode={70563}, 
            state={Baden-Württemberg},
            country={Germany}}

\affiliation[3]{organization={Control and Cyber-Physical Systems, Technical University of Darmstadt},
            addressline={Landgraf-Georg-Straße 4}, 
            city={Darmstadt},
            postcode={64283}, 
            state={Hessen},
            country={Germany}}

\title{Experimentally implemented dynamic optogenetic optimization of {ATPase} expression using knowledge-based and Gaussian-process-supported models}

\begin{abstract}
{Optogenetic modulation of adenosine triphosphatase (ATPase) expression represents a novel approach to maximize bioprocess efficiency by leveraging enforced adenosine triphosphate (ATP) turnover. In this study, we experimentally implement a model-based open-loop optimization scheme for optogenetic modulation of the expression of ATPase. Increasing the intracellular concentration of ATPase, and thus the level of ATP turnover, in bioprocesses with product synthesis coupled with ATP generation, can lead to increased product formation and substrate uptake. Previous simulation studies formulated optimal control problems using dynamic constraint-based models to find optimal light inputs in fermentations with optogenetically mediated ATPase expression. However, using these models poses challenges due to resulting bilevel optimizations and complex parameterization. Here, we outline a simplified unsegregated and quasi-unstructured kinetic modeling approach that reduces the number of dynamic states and leads to single-level optimizations. The models can be augmented with Gaussian processes to compensate for model uncertainties. We implement optimal control constrained by knowledge-based and hybrid models for optogenetic ATPase expression in \textit{Escherichia coli} with lactate as the main product. To do so, we genetically engineer \textit{E. coli} to obtain optogenetic expression of ATPase using the CcaS/CcaR system. This represents the first experimental implementation of model-based optimization of ATPase expression in bioprocesses.} 
\end{abstract}

\begin{graphicalabstract}
\centering
\includegraphics[scale=0.6]{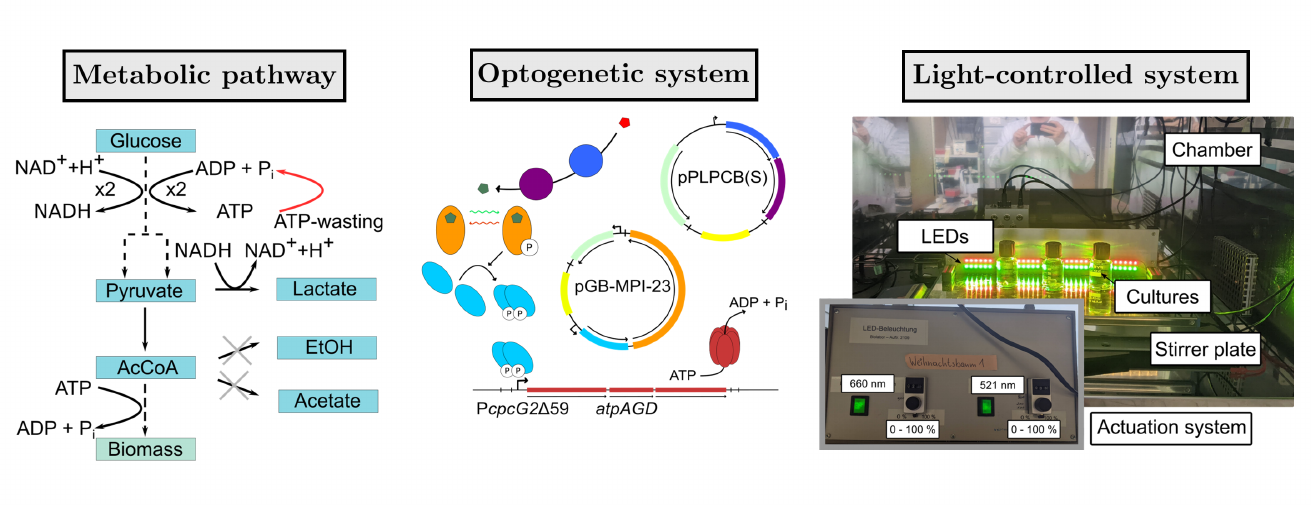}
\end{graphicalabstract}

\begin{highlights}
\item Experimental validation of model-based open-loop optimization of optogenetic ATPase expression in \textit{E. coli}.
\item Simple macro-kinetic modeling can be augmented with Gaussian processes to address model uncertainties.
\item Adjustable batch-to-batch fermentation performance via optogenetic modulation of ATPase.
\item Validated optimization framework paves the way for advanced metabolic engineering applications.
\item Hybrid model-based optimization strategy with the potential for future metabolic cybergenetic implementations.
\end{highlights}


\end{frontmatter}

\section{Introduction}
\label{sec:introduction}
Global challenges such as climate change, depletion of non-renewable fossil resources, and a growing population are driving the search for sustainable production systems. Microbial cell factories are (engineered) microorganisms capable of synthesizing valuable metabolites from renewable resources. They show potential to substitute, e.g., petrochemical production of chemicals, materials, and fuels with biobased and sustainable alternatives \cite{yang_systems_2017,cho_designing_2022}. It is often necessary to optimize both the production processes as well as the cell´s metabolism to achieve product titers, yields, and volumetric productivities that ensure profitability \cite{woodley_towards_2020,lee_factors_2022}. The product yield determines how much substrate is needed to produce a certain amount of product. Volumetric productivity is the rate of product formation per culture volume and determines how fast production occurs.

In \textit{traditional} genetic and metabolic engineering, the cell is rewired to optimize the \textit{steady-state} metabolic flux distribution, often toward maximizing the product yield under a specific cultivation environment. Maximizing the product yield diverges resources from biomass synthesis, thus frequently decreasing the volumetric productivity \cite{banerjee_perspectives_2023}. Furthermore, the gene expression of enzymes involved in production pathways is constitutive in several cases, i.e., always active and happening at a constant rate \cite{holtz_engineering_2010,zhang_design_2012,brockman_dynamic_2015_b}. Thus, from a control engineering perspective, traditional metabolic engineering follows a \textit{static} approach. That is, the cell metabolism is not \textit{a priori} engineered to actively respond to external or internal signals for steering gene expression and metabolism toward desired production \textit{modes}. Of course, the cell´s metabolism naturally reacts to several external/internal signals; this is vital for microorganisms. However, in \textit{static} metabolic engineering, these signals are not dynamically exploited for the optimization of metabolism. Since the cell has a \textit{static} flux distribution for engineered pathways, these pathways cannot adapt to changing conditions, hence lacking operational flexibility. 

Alternatively, cells can be engineered to express metabolism-relevant proteins such as enzymes in an \textit{inducible} and \textit{dynamic} fashion, following a dynamic metabolic engineering approach \cite{brockman_dynamic_2015_b,venayak_engineering_2015,lalwani_current_2018,hartline_dynamic_2021,burg_large-scale_2016}. This idea can be exploited for manipulating metabolic fluxes using external inputs for online process optimization and control \cite{espinel_opt_2022,espinel_cyb_fram_2023}. Dynamic metabolic control can enable, e.g., optimal shift between growth and production metabolic \textit{modes}. It can also help to minimize the metabolic burden associated with the constitutive expression of metabolic enzymes/pathways. Dynamic adenosine triphosphate (ATP) turnover is a promising metabolic control strategy. In bioprocesses where the product synthesis is stoichiometrically coupled with ATP formation, enforcement of ATP loss can lead to an increase in product formation and substrate uptake (cf. e.g. \cite{hadicke_enforced_2015,espinel_ATP_w_2022,espinel_cyb_fram_2023,wichmann_characterizing_2023,boecker_broadening_2019,boecker_increasing_2021,zahoor_atpase-based_2020}).

In previous \textit{simulation-based} studies \cite{espinel_opt_2022,espinel_cyb_fram_2023}, we proposed to put the F\textsubscript{1}-subunit of the adenosine triphosphatase (ATPase) enzyme, responsible for the hydrolysis of ATP into ADP, under the regulation of an optogenetic gene expression system, i.e., making it inducible by light. This would enable one to influence the intracellular amount of ATPase by manipulating light and thereby the level of ATP \textit{wasting}. In this work, we interchangeably use the term \say{ATPase} to refer to the \say{F\textsubscript{1}-subunit} of that enzyme complex. Light can actuate on biological systems in a precise spatiotemporal, orthogonal, and reversible way, which is convenient for external control of gene expression \cite{pouzet_promise_2020,baumschlager_synthetic_2021}. Finding the optimal light trajectories for process optimization imposes, however, several challenges. For instance, for dynamic ATP turnover applications, one needs to carefully fine-tune the level of the F\textsubscript{1}-ATPase, and thus the ATP turnover, to avoid driving the cell into unstable states \cite{klamt_when_2018}.

In \cite{espinel_opt_2022,espinel_cyb_fram_2023}, we showed a \textit{model-based} optimal control problem to find the optimal light trajectories for enhanced product yield via optogenetic modulation of the ATPase. To model the system, we outlined a dynamic constraint-based model that integrates the dynamics of metabolic reactions, the light-inducible genetic actuator, and resource allocation phenomena. The model considers extracellular and intracellular metabolites, as well as intracellular biomass components. The extracellular metabolites and the intracellular biomass components are the dynamic states, while the intracellular metabolites are assumed in a quasi-steady state. The latter model is formulated as an optimization problem subject to constraints and can be considered an extended version of dynamic enzyme-cost flux balance analysis \cite{waldherr_dynamic_2015}.

Although the model in \cite{espinel_opt_2022,espinel_cyb_fram_2023} offers deep insight into metabolism and resource allocation, it can be \textit{too complex} for experimental implementations. It can be technically challenging to gather measurements of the dynamic states to parameterize and validate such a model due to, e.g., a lack of adequate sensors and the unavailability of analytical technologies. Although one could in principle use \textit{soft} sensors (cf. e.g. \cite{jabarivelisdeh_adaptive_2020,espinel_fie_2022}), their suitability depends on the number of states that can be measured and the quality of the underlying mathematical models, which are sometimes limited.

The modeling and optimization framework in \cite{espinel_opt_2022,espinel_cyb_fram_2023} is also computationally expensive. Because the underlying dynamic model is formulated as an optimization problem, the resulting model-based optimal control problem turns out to be a \textit{bilevel optimization}. Solving bilevel optimizations is not trivial. One often needs to make assumptions on the relation between the upper- and lower-level optimization problems, e.g., an \textit{optimistic} or \textit{pessimistic} relation (cf. \cite{dempe_solution_2019,dempe_bilevel_2020,alma99123531260105763} for more details). As done in \cite{espinel_opt_2022,espinel_cyb_fram_2023}, we can reformulate the bilevel optimization into a single-level optimization, following an optimistic approach, by substituting the lower-level problem by its corresponding Karush–Kuhn–Tucker conditions. This results in a mathematical program with complementarity constraints which becomes non-convex given the non-linearity of the complementarity constraints, hence, in general, difficult to solve. The Lagrange multiplier or dual variables, coming from the Karush–Kuhn–Tucker conditions, further increase the size of the optimization problem as they become additional optimization variables.

In this paper, we extend our previous work \cite{espinel_opt_2022,espinel_cyb_fram_2023} by proposing a more straightforward modeling approach for fermentations with optogenetic control of the ATPase. We seek to minimize the number of dynamic states, without sacrificing the model predictability for model-based optimal control of metabolism. To this end, we model only the most relevant extracellular states and the optogenetically controlled intracellular enzyme. In case of significant model uncertainty, we outline the use of Gaussian processes \cite{rasmussen_gaussian_2006}, a machine-learning method, to \textit{learn} the \textit{error} of the \textit{a priori} \textit{known} dynamic equations, thus rendering a \textit{hybrid} model. 

We formulate a single-level model-based dynamic optimization problem using the \textit{simplified} system model. This circumvents bilevel optimization schemes, facilitating the numerics and computational effort. As a major contribution, we experimentally validate our modeling and optimization framework using the anaerobic lactate fermentation of an engineered \textit{Escherichia coli} with optogenetic control of the ATPase, the same biological system considered in \cite{espinel_opt_2022,espinel_cyb_fram_2023}. Remark that the latter references are simulation-based studies, while in this work also experimental validation is shown.

\begin{figure} [htb!]
\begin{center}
\includegraphics[scale=0.38]{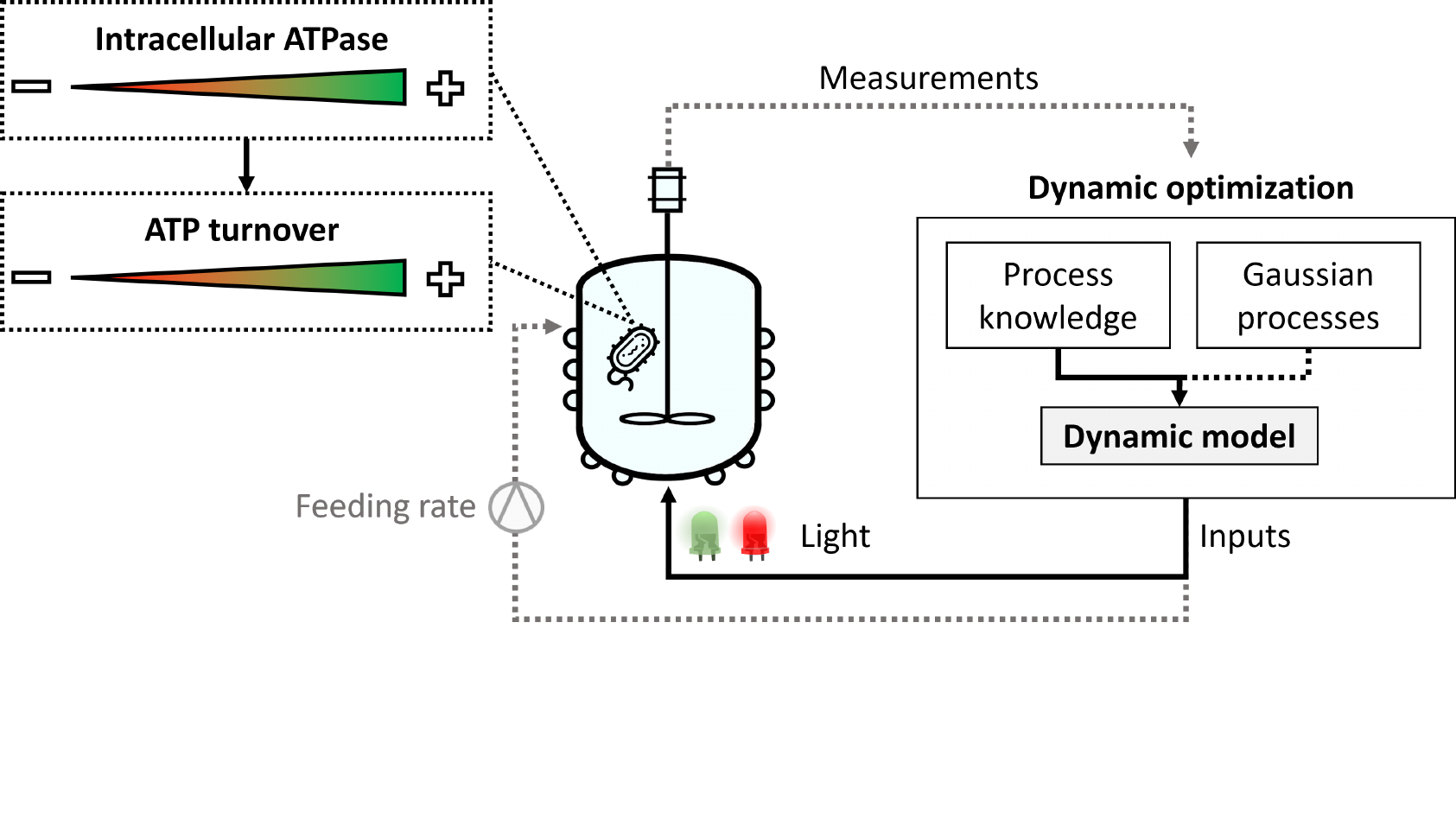}
\vspace{-1.5cm}
\end{center}
\caption{Open-loop control of ATP turnover via optogenetic modulation of the ATPase in batch processes. In gray we show other potential configurations such as closed-loop control and fed-batch fermentations, although these fall out of the scope of this study.}
\label{fig:overview_open_loop}
\end{figure}

A scheme of the dynamic optimization control strategy is presented in Fig.  \ref{fig:overview_open_loop}. Note that the control strategy could be in principle performed both in an open-loop or closed-loop manner. For simplicity, this study focuses exclusively on open-loop control. The process inputs can in principle encompass both \textit{intracellular} and \textit{extracellular} degrees of freedom, e.g., light intensity to modulate ATPase expression (\textit{intracellular}) and substrate feeding rate in case of fed-batch fermentations (\textit{extracellular}). This study specifically focuses on batch processes, thus no feeding rate is considered. Additionally, the model utilized for constraining the dynamic optimization can be solely based on knowledge or of a hybrid nature incorporating machine-learning parts such as Gaussian processes. In this study, we consider both of these model alternatives.

The remainder of this paper is structured as follows. First, we offer an overview of the biological system and optogenetic setup, which will be the basis for introducing our modeling framework and dynamic optimization problem. Afterward, we outline specific model assumptions for the lactate fermentation case with optogenetic control of the ATPase. Finally, we present the experimental results that support the proposed modeling and optimization framework.

\section{Materials and methods}

\subsection{Overview of the biological system and experimental setup}
\label{sec:overview_case}
\begin{figure} [htb!]
\begin{center}
\includegraphics[scale=0.38]{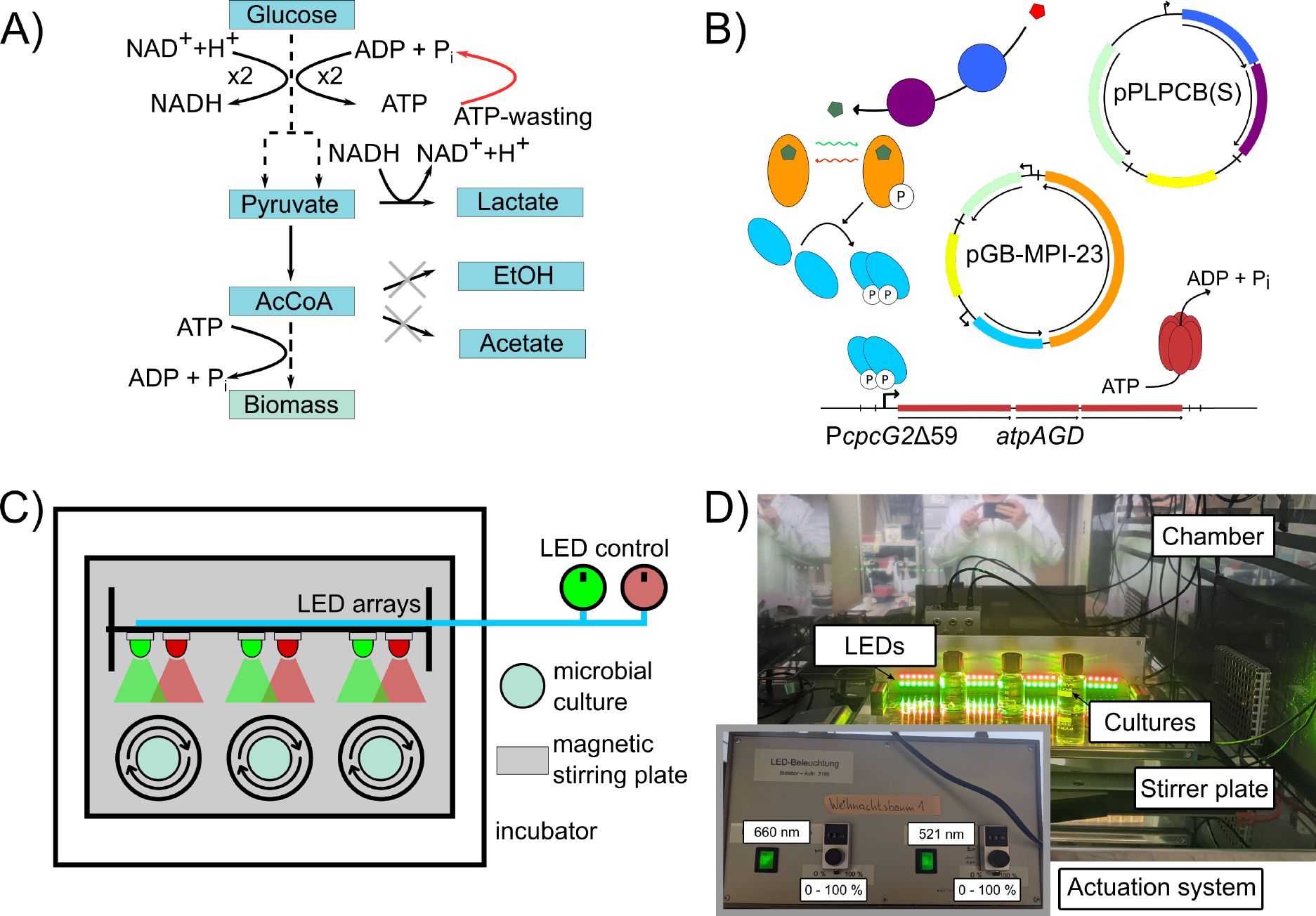}
\end{center}
\caption{A) Simplified core metabolism of the microorganism used in this study, i.e., \textit{E. coli} sGB015 with enforced ATP wasting. It shows the conversion of glucose into lactate. Relevant redox and energy co-factors are presented.
B) The light-inducible ATP wasting is managed by three heterologous genetic elements in \textit{E. coli} sGB015: pPLPCB(S), pGB-MPI-23 and the chromosomal insertion of PcpcG2$\Delta$59-\textit{atpAGD}-rrnBT1. Plasmids are shown circular, while the chromosome is shown linear. Note that genes (italicized) and their related proteins (non-italicized) are shown with the same color. pPLPCB(S) expresses \textit{ho1} (dark blue) and \textit{pcyA} (purple), thereby enabling the conversion of heme (red pentagon) into phycocyanobilin (green pentagon), the chromophore necessary for CcaS to detect light. The expression of \textit{ccaS} (orange) and \textit{ccaR} (pale blue) is enabled through pGB-MPI-23. CcaS autophosphorylates after a conformational change induced by green light with the photon-protein interaction enabled through phycocyanobilin.
Afterward, CcaS phosphorylates CcaR (phosphate group P is represented by circles), leading to CcaR dimerization and functioning as a transcription factor for PcpcG2$\Delta$59 on the chromosome, thereby initiating the expression of \textit{atpAGD} (red). Promoters are presented as arrows on the plasmids and the chromosome. 
Open reading frames are shown as arrows next to their related genes. Terminators are marked as black perpendicular lines. Origins of replication are shown in yellow and antibiotic resistances in pale green.
C) Scheme of the fermentation setup with the green and red light delivery system based on LEDs.
D) Photograph of the actual setup shown in C). The temperature regulation chamber, magnetic stirrer plate, LED arrays, and actuation system are labeled. The actuation system is shown with the two tunable regulators for determining the current that is delivered to the green and red LED arrays.
}
\label{fig:overview}
\end{figure}

Under anaerobic fermentation of glucose, lactate production is net ATP positive. As a basis to showcase the proposed model-based optimization strategy of the optogenetically regulated ATPase, we consider an \textit{E. coli} strain with blocked ethanol and acetate pathways (cf. Fig.\ref{fig:overview}-A) \cite{hadicke_enforced_2015}. Therefore, lactate synthesis becomes the main fermentation pathway to achieve redox balance, making lactate production suitable for enforced ATP turnover \cite{hadicke_enforced_2015,espinel_ATP_w_2022,espinel_cyb_fram_2023,wichmann_characterizing_2023}.

We aim to control gene expression of the ATPase using light as an external input. To do so, we utilize the two-component system known as CcaS/CcaR (chromatic acclimation sensor/regulator) to establish control over ATPase F\textsubscript{1}-subunit (\textit{atpAGD}) expression in \textit{E. coli}. The CcaS/CcaR system, originated from cyanobacteria, allows for the regulation of gene expression by changing the red-to-green light ratio to which the cells are exposed (cf. e.g. \cite{milias-argeitis_automated_2016,olson_characterizing_2014,senoo_lightinducible_2019}). Green light causes CcaS to autophosphorylate and then phosphorylate CcaR. CcaR dimerizes when phosphorylated, becoming an activating transcription factor. Red light causes CcaS to dephosphorylate and gene expression is repressed.

To construct our biological system, we inserted an \textit{atpAGD} expression cassette into the chromosome of \textit{E. coli} in which all genes are regulated by the optimized promoter of \textit{cpcG2} \cite{schmidl_refactoring_2014}. Subsequently, we introduced plasmid pPLPCB(S) for the production of phycocyanobilin, a cofactor necessary for photo-sensing, as well as plasmid pGB-MPI-23 to express the CcaS/CcaR proteins. This resulted in the final strain \textit{E. coli} sGB015 where ATPase expression is regulated by green and red light (cf. Fig. \ref{fig:overview}-B). Refer to Section \ref{sec:genetic_procedure} for the detailed genetic engineering procedure.

Anaerobic fermentation experiments with \textit{E. coli} sGB015 subjected to optogenetic manipulation were carried out at 37 \textdegree C in a Certomat BS-1 incubator (B. Braun Biotech International) (cf. Fig. \ref{fig:overview}-C). Red and green light outputs were generated through light-emitting diode (LED) arrays (Osram OSLON SSL LED green $\lambda_{\textnormal{peak}}$ = 521 nm; Osram OSLON SSL LED red $\lambda_{\textnormal{peak}}$ = 660 nm), whereby the emitted light was regulated via the supplied current. The corresponding photon flux density in $\mu$mol$\ $m$^{-2}\ $s$^{-1}$ units was determined using an ULM-500 Universal Light Meter (Heinz Walz GmbH). All experiments were performed using three biological replicates. For more details on the fermentation procedure and analytical measurements, refer to Section \ref{sec:fermentation_procedure}.

\subsubsection{Genetic engineering procedure}
\label{sec:genetic_procedure}
The strain KBM10111s ($=$MG1655 $\Delta$\textit{adhE} $\Delta$\textit{ackA}-\textit{pta}) \cite{hadicke_enforced_2015} was used as a basis to create an \textit{E. coli} strain that produces lactate as the main fermentation product and portrays a light-inducible expression of \textit{atpAGD}. First, pGB-MPI-035, a modified version of pSKA397 \cite{milias-argeitis_automated_2016}, was transformed into KBM10111s together with pTNS3 \cite{Choi2008} for Tn7-based insertion of \textit{atpAGD} regulated by PcpcG2$\Delta$59 \cite{schmidl_refactoring_2014}, downstream of \textit{glmS} \cite{Choi2005}. 
This resulted in sGB013 ($=$MG1655 $\Delta$\textit{adhE} $\Delta$\textit{ackA}-\textit{pta} Tn7::\textit{cat}-PcpcG2$\Delta$59-\textit{atpAGD}-rrnBT1). The chloramphenicol resistance cassette of this strain was removed though FLP activity by transformation with pCP20 \cite{Cherepanov1995}, resulting in sGB014 ($=$MG1655 $\Delta$\textit{adhE} $\Delta$\textit{ackA}-\textit{pta} Tn7::PcpcG2$\Delta$59-\textit{atpAGD}-rrnBT1). 
The genes for the light-regulated transcription system, \textit{ccaS}/\textit{ccaR} controlled through PccaR, were introduced in a \textit{sfgfp}-deficient variant of pSKA413 \cite{milias-argeitis_automated_2016} (pGB-
MPI-23). Furthermore, the \textit{Synechocystis} PCC6803 genes \textit{ho1} and \textit{pcyA}, enabling the conversion of heme into phycocyanobilin, were introduced by transformation with pPLPCB(S) \cite{tabor_multichromatic_2011}. Both plasmids were transformed into sGB014, resulting in the final strain used for all experiments, i.e., \textit{E. coli} sGB015 ($=$MG1655 $\Delta$\textit{adhE} $\Delta$\textit{ackA}-\textit{pta} Tn7::PcpcG2$\Delta$59-\textit{atpAGD}-rrnBT1 pPLPCB(S) pGB-MPI-23). Sequence files for all genetic elements created in this work can be found in genebank format through the Edmond repository: \href{https://doi.org/10.17617/3.H5GT8I}{https://doi.org/10.17617/3.H5GT8I}.

\subsubsection{Fermentation experiments and analytical measurements}
\label{sec:fermentation_procedure}
Single colonies of \textit{E. coli} sGB015 were used to start 10 ml aerobic cultures with LB$_{\textnormal{0}}$ (10 g/l tryptone, 5 g/l yeast extract, 5 g/l NaCl) in 100 ml shake flasks with baffles at 37 \textdegree C and 200 rpm. Next, precultures with standard defined medium 
(4 g/l glucose, 
34 mM NaH$_{\textnormal{2}}$PO$_{\textnormal{4}}$, 
64 mM K$_{\textnormal{2}}$HPO$_{\textnormal{4}}$, 
20 mM (NH$_{\textnormal{4}}$)$_{\textnormal{2}}$SO$_{\textnormal{4}}$, 
9.52 mM NaHCO$_{\textnormal{3}}$, 
1 $\mu$M Fe(SO$_{\textnormal{4}}$)$_{\textnormal{4}}$, 
300 $\mu$M MgSO$_{\textnormal{4}}$, 
1 $\mu$M ZnCl$_{\textnormal{2}}$, 
10 $\mu$M CaCl$_{\textnormal{2}}$) \cite{Tanaka1967} with 150 $\mu$g/ml spectinomycin and 25 $\mu$g/ml chloramphenicol were inoculated and grown overnight under red light in 50 ml Schott flasks with 25 ml culture volume at 37 \textdegree C and 180 rpm. The main fermentation experiments were inoculated from the latter preculture in fresh standard defined medium, cultivated in 50 ml flasks with 50 ml culture volume. For sampling purposes, the vessels were briefly transferred inside a Whitley A25 anaerobic workstation (Meintrup DWS Laborgeräte GmbH) with an oxygen-free atmosphere (80 \% N$_{\textnormal{2}}$, 10 \% CO$_{\textnormal{2}}$, 10 \% H$_{\textnormal{2}}$). During cultivation, the flasks were tightly closed to prevent gas exchange.

Glucose concentrations were measured with the HK assay kit (Megazyme Ltd.). 
Lactate was measured by reversed phase HPLC utilizing an Inertsil ODS-3 column (5 $\mu$m, RP-18 100A, 250 x 4.6 mm)
(GL Sciences Inc.) with a flow rate of 1.0 ml/min, using a running buffer consisting of 0.1 M NH$_4$H$_2$PO$_4$ at pH 2.6 and 40 \textdegree C. 
The injection volume was 10 $\mu$l and detection was performed with a UV-DAD detector at 210 nm.

\subsection{Modeling of fermentations with optogenetic ATPase expression}
\label{sec:hybr_modeling}
We outline the general structure of our modeling approach for fermentations with optogenetic regulation of the ATPase expression. 
For clarity of presentation, we keep for the moment the notation general, while in Section \ref{sec:lactate_model} we elaborate on the specific model assumptions for the lactate fermentation case study. Note that we use bold fonts for vectors and matrices, and non-bold fonts for scalar variables and parameters.

\subsubsection{General model formulation}
We consider the dynamics of biomass $B\in \mathbb{R}$, \textit{rate-limiting} external substrates $\bm{s}\in \mathbb{R}^{n_s}$, \textit{rate-limiting} (by)products and products of interest $\bm{p} \in \mathbb{R}^{n_p}$, as well as the intracellular ATPase $E \in \mathbb{R}$. While $B$, $\bm{s}$, and $\bm{p}$ are expressed in mass per culture volume, $E$ is expressed in mass of ATPase per mass of cells. We treat the biomass as a homogeneous population of cells, hence the model is \textit{unsegregated} and the biomass is modeled as a single component. Since we lump up intracellular metabolism, except for the dynamics of the ATPase expression, the model is also \textit{quasi-unstructured}. The dynamic input $u \in \mathbb{R}$ is the green light photon flux density. Based on these considerations, the process dynamics read
\begin{subequations} 
\begin{align}
&\odv{B(t)}{t} =  \bm{S_{B}}\bm{r}(\bm{x}(t),u(t),\bm{\theta}) + \bm{Q_B}\bm{w}(\bm{x}(t),u(t),\bm{\tau}),\label{eq:ode_B_general}\\
&\odv{E(t)}{t} =  \bm{S_{E}}\bm{r}(\bm{x}(t),u(t),\bm{\theta}) + \bm{Q_E}\bm{w}(\bm{x}(t),u(t),\bm{\tau}),\label{eq:ode_E_general}\\
&\odv{\bm{s}(t)}{t} = \bm{S_{s}}\bm{r}(\bm{x}(t),u(t),\bm{\theta}) + \bm{Q_s}\bm{w}(\bm{x}(t),u(t),\bm{\tau}),\label{eq:ode_s_general} \\
&\odv{\bm{p}(t)}{t} = \bm{S_{p}}\bm{r}(\bm{x}(t),u(t),\bm{\theta}) + \bm{Q_p}\bm{w}(\bm{x}(t),u(t),\bm{\tau}),\label{eq:ode_p_general} \\
&\bm{x}:=[B,E,\bm{s}^\tran,\bm{p}^\tran]^\tran,\label{eq:x_def} \\
&\bm{x}(t_0)=\bm{x_0},\label{eq:x0}
\end{align}
\end{subequations}
where $t\in [t_0,t_f]\subset \mathbb{R}_{\geq 0}$ is the process time, $t_0$ the initial process time, and $t_f$ the final process time.

In the previous equations, $\bm{r}: \mathbb{R}^{n_x} \times \mathbb{R} \times \mathbb{R}^{n_\theta} \rightarrow \mathbb{R}^{n_r}$ is a vector-valued function that comprises the reaction rates of the process, e.g., production, consumption, degradation, and dilution rates. $\bm{\theta}\in \mathbb{R}^{n_\theta}$ are parameters of the reaction rates. $\bm{S_{B}} \in \mathbb{R}^{1 \times n_r}$, $\bm{S_{E}} \in \mathbb{R}^{1 \times n_r}$, $\bm{S_{s}} \in \mathbb{R}^{n_s \times n_r}$, $\bm{S_{p}} \in \mathbb{R}^{n_p \times n_r}$ map the coefficients of the reaction rates to the differential equations. The previous terms comprise the \textit{knowledge-based} part of the model. Note that modeling $E$ allows one to capture possible time delays in the extracellular rates arising from the lumped transcription/translation dynamics of the ATPase. Furthermore, we consider in the dynamic equations model uncertainty due to, e.g., oversimplified or wrong model assumptions resulting from the lack of mechanistic description of the intracellular metabolism. The model \textit{error} is
defined by the vector-valued function $\bm{w}: \mathbb{R}^{n_x} \times \mathbb{R} \times \mathbb{R}^{n_\tau} \rightarrow \mathbb{R}^{n_w}$ and it aims at capturing the dynamics neglected or misrepresented by the knowledge-based part of the dynamic equations. $\bm{\tau}\in \mathbb{R}^{n_\tau}$ comprises the parameters of $\bm{w}$. $\bm{Q_{B}} \in \mathbb{R}^{1 \times n_w}$, $\bm{Q_{E}} \in \mathbb{R}^{1 \times n_w}$, $\bm{Q_{s}} \in \mathbb{R}^{n_s \times n_w}$, $\bm{Q_{p}} \in \mathbb{R}^{n_p \times n_w}$ map the functions describing the model error to the corresponding differential equations. Hereafter, we will omit the time dependency of the variables when clear from the context.

Naturally, if the knowledge-based part of the dynamic equations describes the real system sufficiently well, i.e., without considering the model error $\bm{w}$, one can simply neglect $\bm{w}$ from the final model. However, in cases of significant model-plant mismatch, we propose to create \textit{hybrid} models where we \textit{learn} $\bm{w}$ from process data with machine learning. Experiments in biotechnology are, nevertheless, often expensive and time-consuming, thus large and high-quality training data sets for machine learning are typically scarce. Here, we focus on Gaussian processes \cite{rasmussen_gaussian_2006}, a machine-learning method that can in principle offer good predictability even if trained with small data sets.

\subsubsection{Gaussian processes}
\label{sec:gp_regression}
Gaussian processes are classified as \textit{probabilistic} machine-learning methods, whereby the predictions contain a measurement of the prediction \textit{uncertainty}. Gaussian processes render a probabilistic \textit{distribution} over functions, as opposed to other approaches such as conventional neural networks where the predictions are deterministic. In this section, we present a general overview of Gaussian processes; we refer the reader to \cite{rasmussen_gaussian_2006,pohlodek_flexible_2022,morabito_efficient_2022,himmel_machine_2023} for more information.

Let $l \in \mathbb{R}$ be the label (regression output) of one Gaussian process regressor and $\bm{v} \in \mathbb{R}^{n_v}$ the corresponding features (regression inputs). The Gaussian process regressor aims to model an unknown function $h: \mathbb{R}^{n_v} \rightarrow \mathbb{R}$ using
noisy observations $l$ of $h(\bm{v})$
\begin{equation}
    l = h(\bm{v})+\epsilon,
\end{equation}
where $\epsilon$ is Gaussian distributed \textit{measurement} noise $\epsilon \sim \mathcal{N}(0,\sigma_n^2)$ with zero mean and variance $\sigma_n^2$. 

Let us define $\bm{V} \in \mathbb{R}^{n_v \times n_d}$ as the matrix of the supplied training inputs and $\bm{L} \in \mathbb{R}^{1 \times n_d}$ as the training outputs, where $n_d$ is the number of training data sets. Furthermore, let $\bm{v_i} \in \mathbb{R}^{n_v \times 1}$ and $\bm{v_j} \in \mathbb{R}^{n_v \times 1}$ be two arbitrary input vectors. 

In Gaussian processes, we assume that the labels are normally distributed
\begin{equation}
    h(\bm{v}) \sim \mathcal{N}(m(\bm{v}),k(\bm{v},\bm{v})),
\end{equation}
where $m:\mathbb{R}^{n_v} \rightarrow \mathbb{R}$ is the mean \textit{function} and $k: \mathbb{R}^{n_v} \times \mathbb{R}^{n_v} \rightarrow \mathbb{R}$ is the kernel or covariance \textit{function}. Overall, the kernel function describes the \textit{neighborhood} or \textit{similarity} between data points in the feature space. 

Gaussian processes start from a \textit{prior distribution} of functions, characterized by a prior mean function and a prior covariance function, i.e., \textit{prior} to observing data. The prior distribution is determined by the choice of the kernel function, which is chosen to be infinitely differentiable, smooth, and continuous. Many kernel functions are possible; in this work, we use the squared-exponential kernel function
\begin{equation}
    k(\bm{v_i},\bm{v_j}|\bm{\tau}) = \sigma^2 \mathrm{exp}\left( \frac{-(\bm{v_i}-\bm{v_j})^\tran(\bm{v_i}-\bm{v_j})}{2d^2}\right),
\end{equation}
where $\sigma^2 \in \mathbb{R}$ is the signal variance, $d\in \mathbb{R}$ is the length-scale, both of which are \textit{hyperparameters} of the kernel function.

Furthermore, we obtain the covariance \textit{matrix} $\bm{K} \in \mathbb{R}^{n_d \times n_d}$ based on the chosen kernel function and the supplied training data
\begin{equation}
\bm{K}=\begin{bmatrix}
  k(\bm{v_1},\bm{v_1})  & \cdots  & k(\bm{v_1},\bm{v_{n_d}}) \\ 
\vdots & \ddots & \vdots\\ 
 k(\bm{v_{n_d}},\bm{v_{1}}) & \cdots &  k(\bm{v_{n_d}},\bm{v_{n_d}}) 
\end{bmatrix},
\end{equation}
which captures the relationship between the features.

We optimize the hyperparameters of the kernel function by maximizing the log marginal likelihood, i.e., $\bm{\tau}^* = \mathrm{arg} \, \mathrm{max}_{\bm{\tau}} \log p(\bm{L}|\bm{V},\bm{\tau})$, with
\begin{align}
    &\log(p(\bm{L}|\bm{V},\bm{\tau}))=-\frac{1}{2}\bm{L}^T(\bm{K}+\sigma_\mathrm{n}^2I)^{-1}\bm{L} -\frac{1}{2}\mathrm{log}\left(|\bm{K}+\sigma_n^2\bm{I}|\right) - \frac{n_d}{2}\mathrm{log}(2\pi),
\end{align}
where $\bm{\tau}:=[\sigma^2,d,\sigma_n^2]$ and $\bm{I}$ is the identity matrix of appropriate size.

Finally, the conditional posterior of the Gaussian process with optimal hyperparameters follows a normal distribution for a test input vector $\bm{v^*} \in \mathbb{R}^{n_v \times 1}$, i.e., $p(\bar{h}(\bm{v}^*)|\bm{V},\bm{L}) \sim \mathcal{N}(\bar{h},\Sigma)$, with predictive mean $\bar{h}$ and variance $\Sigma$
\begin{subequations} 
\begin{align}
&\bar{h}(\bm{v}^*)=\bm{\Tilde{k}}^\tran(\bm{K}+\sigma_n^2\bm{I})^{-1}\bm{L}, \\
&\Sigma(\bm{v}^*) = k(\bm{v}^*,\bm{v}^*) - \bm{\Tilde{k}}^\tran(\bm{K}+\sigma_n^2\bm{I})^{-1}\bm{\Tilde{k}}, \\
&\bm{\Tilde{k}} := [k(\bm{v_1},\bm{v}^*),...,k(\bm{v_{n_d}},\bm{v}^*)]^\tran.
\end{align}
\end{subequations}

Since we aim to capture with Gaussian processes the model-plant mismatch $\bm{w}$ of differential equations, then $h(\bm{v}):=w_i(\bm{v})$, where $w_i$ is the model error of a differential equation $i$. We consider as many Gaussian process regressors as $n_w$, i.e., multi-input single-output Gaussian processes. The features of the Gaussian process can be, e.g., appropriate model states and inputs. 

\subsection{Open-loop dynamic optimization of the optogenetically modulated ATPase expression}
\label{sec:olo_framework}
We propose to find the optimal light input trajectories that maximize the efficiency of the process with optogenetic control of the ATPase by solving the following optimal control problem
\label{sec:opt_control}
\begin{maxi!} 
    {u(\cdot),\bm{x_0}}{J(\cdot),\label{eq:optimal_cost}}{\label{eq:optimal}}{}
    \addConstraint{}{\text{Eqs.}\eqref{eq:ode_B_general}-\eqref{eq:x0},}{}
    \addConstraint{}{0 \leq \bm{g}(\bm{x},u,\bm{\theta},\bm{\tau}), \label{eq:g_cons}}{}
\end{maxi!}
where $J(\cdot)$ is the cost function that captures the efficiency of the process and $\bm{g}: \mathbb{R}^{n_x} \times \mathbb{R} \times \mathbb{R}^{n_\theta} \times \mathbb{R}^{n_\tau} \rightarrow \mathbb{R}^{n_g}$ are additional process constraints. The decision variables of the optimization problem can include in principle the dynamic input, in this case, green light photon flux density, and the initial state concentrations, e.g., initial substrate concentration. In other words, we cover both \textit{static} and \textit{dynamic} degrees of freedom in the formulation above. The cost function in \eqref{eq:optimal_cost} can be the product volumetric productivity, titer, yield, or an explicit function that captures the economic profit of the process. The constraints in Eq. \eqref{eq:g_cons} can include, e.g., safety, economic, or technical constraints.

\subsection{Numerical methods}
The parameter estimation procedure for the knowledge-based part of the model was performed in COPASI using the particle swarm algorithm \cite{hoops_copasicomplex_2006}. We used HILO-MPC \cite{pohlodek_flexible_2022}, a Python toolbox for machine-learning-supported optimal control, for training the Gaussian process regressors and solving the open-loop optimization problems.

\section{Results and discussion}
\subsection{Model of the lactate fermentation with optogenetic control of the ATPase}
\label{sec:lactate_model}
Having described the general modeling and optimization framework in Sections \ref{sec:hybr_modeling} and \ref{sec:olo_framework}, we proceed to describe the proposed model for the lactate fermentation case study in batch mode. We consider the following dynamic states: glucose $s_G \in \mathbb{R}$, lactate $p_L \in \mathbb{R}$, \textit{E. coli}´s biomass $B_c \in \mathbb{R}$, and the intracellular ATPase $E \in \mathbb{R}$. Therefore, in the case study, ${\bm{s}} := s_G$, ${\bm{p}} := p_L$, and $B:=B_c$. Compared to the dynamic constraint-based model proposed in \cite{espinel_opt_2022,espinel_cyb_fram_2023}, this represents a significant reduction in the number of dynamic states, i.e., from 23 to only 4 states. The control input is the green light photon flux density $u_l \in \mathbb{R}$, hence ${\bm{u}} := u_l$.

The proposed model follows
\begin{subequations} 
\begin{align} 
&\odv{s_G}{t} = -q_G(s_G,E,\bm{\theta}) B_c + w_G(s_G,B_{c},p_L,E,u_l,\bm{\tau}),\label{eq:ode_glc} \\ 
&\odv{B_{c}}{t} = \mu(s_G,E,\bm{\theta}) B_{c} + w_c(s_G,B_{c},p_L,E,u_l,\bm{\tau}),\label{eq:ode_bio} \\
&\odv{p_L}{t} = q_L (s_G,E,\bm{\theta}) B_c + w_L(s_G,B_{c},p_L,E,u_l,\bm{\tau}),\label{eq:ode_lac} \\
&\odv{E}{t} = q_E(u_l,\bm{\theta}) - d_E(E,\bm{\theta}),\label{eq:ode_E} \\
&s_G(t_0)=s_{G_0},B_c(t_0)=B_{c_0},p_L(t_0)=p_{L_0},E(t_0)=E_0,
\end{align}
\end{subequations}
where $q_G$, $\mu$, $q_L$, $q_E$, and $d_E$ are \textit{known} kinetic functions with appropriate parameters, and $w_G$, $w_c$, and $w_L$ are Gaussian-process regression functions with appropriate parameters describing the model \textit{error}. In Eqs. \eqref{eq:ode_glc}-\eqref{eq:ode_E}, we assume only two rate-limiting components in the kinetic functions, namely the substrate glucose and the light-inducible intracellular ATPase. We neglect dilution or degradation effects in equations \eqref{eq:ode_glc}-\eqref{eq:ode_lac}. 

Measuring the extracellular concentrations to \textit{parameterize} the model is relatively straightforward, however, measuring the intracellular ATPase concentration remains a challenge. When available and affordable, proteomics \cite{shuken_introduction_2023} can be an option to measure the intracellular ATPase, but it is generally time-consuming and requires dedicated sample preparation. Note that, although proteomics is effective \textit{offline}, implementing it for real-time monitoring remains a challenge. Alternatively, biosensors could, in principle, be used to estimate the intracellular ATPase, for example, by attaching fluorescent protein tags to the enzyme \cite{thorn_genetically_2017}. However, in some cases, this can lead to protein malfunction \cite{weill_assessment_2019}, and the intrinsic folding and maturation of fluorescent proteins \cite{thorn_genetically_2017} may cause delays in the output reading. One could argue that if the ATPase cannot be measured, it may still be estimated with soft sensors as shown in \cite{espinel_fie_2022,espinel_cyb_fram_2023}. Yet, for soft sensors, we require a suitable \textit{validated} mathematical model, which, in the context of this study, is unavailable. For these reasons, we did not implement here a biosensor or a soft sensor for ATPase determination, but it can be the focus of future studies, in particular within feedback control schemes where real-time monitoring is desirable.

Since in our experimental setup we cannot measure the intracellular ATPase, we regard $E$ as a \textit{virtual variable} expressed in virtual units (VU) per gram of biomass and we assume no model uncertainty in Eq. \eqref{eq:ode_E}. For simplicity, we assume that the dynamics of the intracellular ATPase are a function of only the ATPase concentration and the light input (cf. Eq. \eqref{eq:ode_E}). Note that the inoculum/preculture preparation in the fermentations follows a well-standardized protocol (cf. Section \ref{sec:fermentation_procedure}), e.g., the preculture was always grown under red light conditions, i.e., without ATPase induction. Therefore, we arbitrarily set $E(t_0)=0 \, \mathrm{VU/g}$ in all fermentation experiments. We neglect the \textit{exact} biological meaning of Eq. \eqref{eq:ode_E} and associated parameters as long as they help to describe well the dynamics of the extracellular concentrations. As mentioned before, Eq. \eqref{eq:ode_E} allows coupling the input-dependent intracellular state of the cell (micro-scale variable) to the extracellular species (macro-scale variables). By doing so, one can capture time delays in the macro-scale fermentation dynamics arising from the lumped transcription and translation of the intracellular ATPase.

We model the kinetic rates as follows
\begin{subequations} 
\begin{align} 
&q_G(s_\mathrm{G},E) = q_{G_\mathrm{max}}\left( \frac{s_G}{s_G+k_G}\right) \left( 1 + \frac{E^{n_1}}{E^{n_1}+k_\mathrm{GV}^{n_1}} \right),\label{eq:q_G} \\ 
&\mu(s_G,E) = Y_\mathrm{BG}\left( q_G(s_\mathrm{G},E)-m_G \right) \left( 1 - \frac{E^{n_2}}{E^{n_2}+k_\mathrm{BV}^{n_2}} \right),\label{eq:mu} \\
&q_L(s_G,E) = \left(Y_\mathrm{LB}\mu(s_G,E) + m_L\right) \left( 1 + \frac{E^{n_3}}{E^{n_3}+k_\mathrm{LV}^{n_3}} \right),\label{eq:q_L} \\
&q_E(u_l) = q_{E_0} + q_{E_\mathrm{max}} \frac{u_l^{n_4}}{u_l^{n_4}+k_u^{n_4}},\label{eq:q_E} \\
&d_E (E) = k_d E,\label{eq:d_E}
\end{align}
\end{subequations}
where $\bm{\theta}:=[k_\mathrm{BV},k_G,k_\mathrm{GV},k_\mathrm{LV},m_G,m_L,n_1,n_2,n_3,q_{G_\mathrm{max}},Y_\mathrm{BG},Y_\mathrm{LB},q_{E_0},q_{E_\mathrm{max}},n_4,k_u,k_d]^\tran$ comprises the 17 parameters of the assumed \textit{known} kinetic functions. 

If we assume no enforced ATP turnover, i.e., $E(t)=0$, the specific substrate uptake rate (Eq. \eqref{eq:q_G}) follows conventional Monod-type kinetics \cite{heijnen_derivation_1995}, the specific growth rate (Eq. \eqref{eq:mu}) follows the Pirt´s equation for substrate distribution \cite{noauthor_maintenance_1965}, and the specific lactate production rate (Eq. \eqref{eq:q_L}) follows the Leudeking-Piret´s equation for catabolic products \cite{luedeking_kinetic_1959}. We include Hill-type \textit{activation} terms \cite{hill1910possible} in Eqs. \eqref{eq:q_G}-\eqref{eq:q_L} to account for either an increase (+) or decrease (-) in the specific rates as a result of the enforced ATP turnover, i.e., for $E(t) > 0$. The production of the ATPase is \textit{activated} by green light following the Hill function \cite{hill1910possible}, Eq. \eqref{eq:q_E}. Note that we assume homogeneous light penetration in the bioreactor. Finally, we assume an \textit{average} lumped dilution/degradation rate of ATPase based on first-order kinetics, Eq. \eqref{eq:d_E}. 

\subsection{Model validation} \label{sec:validation}
\begin{figure*}[htb!]
\centering
\begin{subfigure}{0.38\textwidth}
  \captionsetup{justification=centering} 
  \caption{Lactate on glucose yield\label{fig:subfig-a}}
  \includegraphics[width=\textwidth, keepaspectratio]{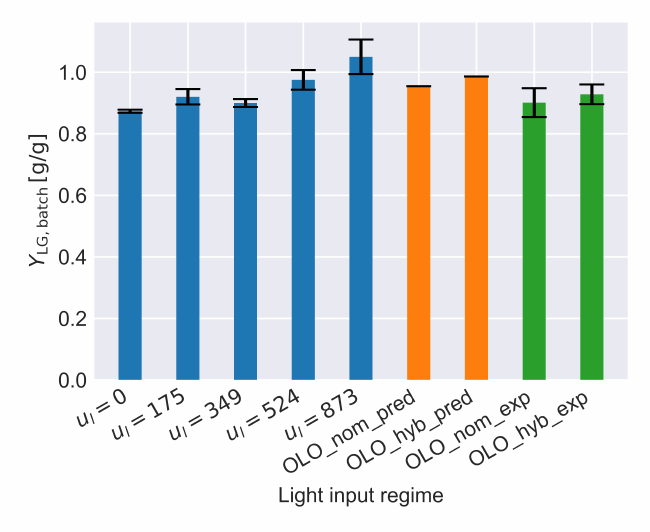}
\end{subfigure}
\vspace{-0.25cm}

\begin{subfigure}{0.38\textwidth}
  \captionsetup{justification=centering} 
  \caption{Biomass on glucose yield\label{fig:subfig-b}}
  \includegraphics[width=\textwidth, keepaspectratio]{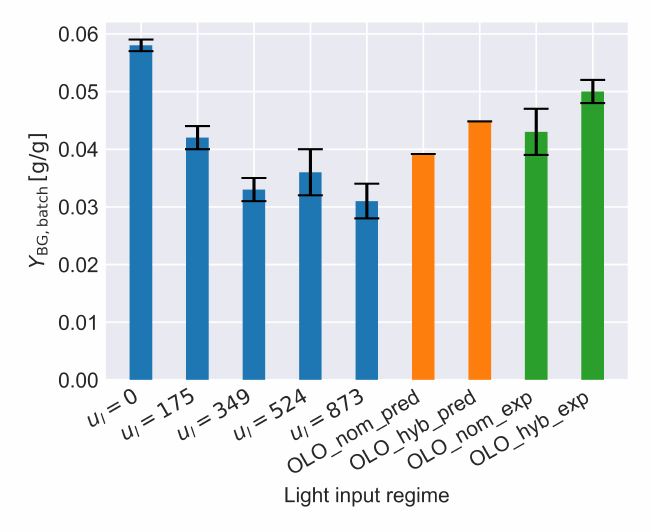}
\end{subfigure}
\vspace{-0.25cm}

\begin{subfigure}{0.38\textwidth}
  \captionsetup{justification=centering} 
  \caption{Lactate volumetric productivity\label{fig:subfig-c}}
  \includegraphics[width=\textwidth, keepaspectratio]{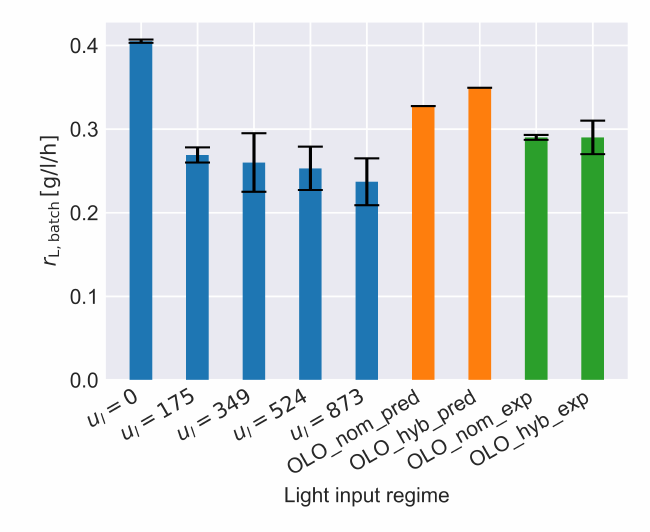}
\end{subfigure}
\caption{A) Average lactate on glucose yield $Y_\mathrm{LG,batch}$, B) biomass on glucose yield $Y_\mathrm{BG,batch}$, and C) lactate volumetric productivity $r_\mathrm{L,batch}$. Blue bars: experiments used to model the system. Orange bars: predicted open-loop optimization results using the nominal (OLO\_nom\_pred) and hybrid (OLO\_hyb\_pred) models. Green bars: actual experimental results for the open-loop optimizations using the nominal (OLO\_nom\_exp) and hybrid (OLO\_hyb\_exp) models. These metrics correspond to average batch results, i.e., considering initial and final points. The units of $u_l$ are $\mu$mol$\ $m$^{-2}\ $s$^{-1}$.}
\label{fig:fermentation_metrics}
\end{figure*}

To validate the proposed model, we ran five fermentation experiments under different constant green light input values, namely 0, 175, 349, 524, and 873 $\mu$mol$\ $m$^{-2}\ $s$^{-1}$. As can be seen in the blue bars in Fig. \ref{fig:fermentation_metrics}, the yield of lactate on glucose for the entire batch $Y_\mathrm{LG,batch}$ increases with increasing light input, up to around 1 g/g, i.e., the maximum theoretical yield. Note that there cannot be biomass generation if the lactate yield is at its maximum theoretical value. Thus, there must have been a slight overestimation of the actual lactate yield, in particular for the experiment with constant $u_l=873$ $\mu$mol$\ $m$^{-2}\ $s$^{-1}$, very likely arising from a systematic measurement error of lactate concentration. Similarly, an increasing light input translates into decreasing biomass on glucose yield $Y_\mathrm{BG,batch}$ and lactate volumetric productivity $r_\mathrm{L,batch}$ averaged over the batch. This is the expected behavior since higher light photon flux density is linked to higher ATPase induction level, thus higher ATPase accumulation and ATP wasting \cite{espinel_opt_2022,espinel_cyb_fram_2023}.

\begin{figure}[]
\centering
\vspace{-0.30cm}
\begin{flushleft} 
\footnotesize{A) Batch with constant $u_l = 0 \, \mathrm{\mu mol \, m^{-2} \, s^{-1}}$}
\end{flushleft} 
\vspace{-0.30cm}
\subfloat{\includegraphics[width=0.28\textwidth, keepaspectratio]{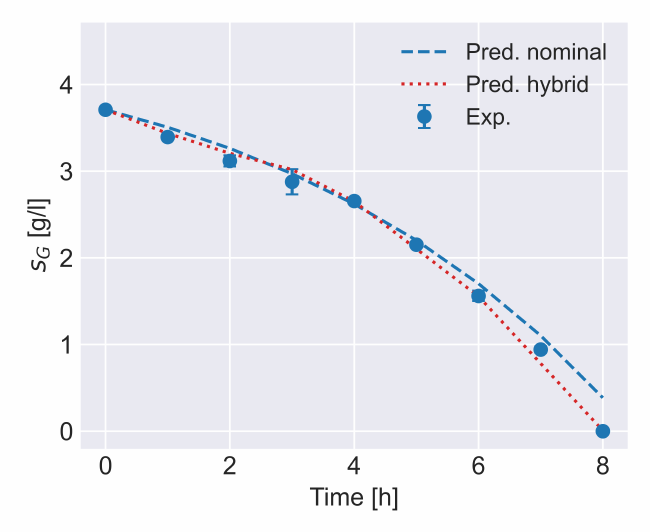}}
\subfloat{\includegraphics[width=0.28\textwidth, keepaspectratio]{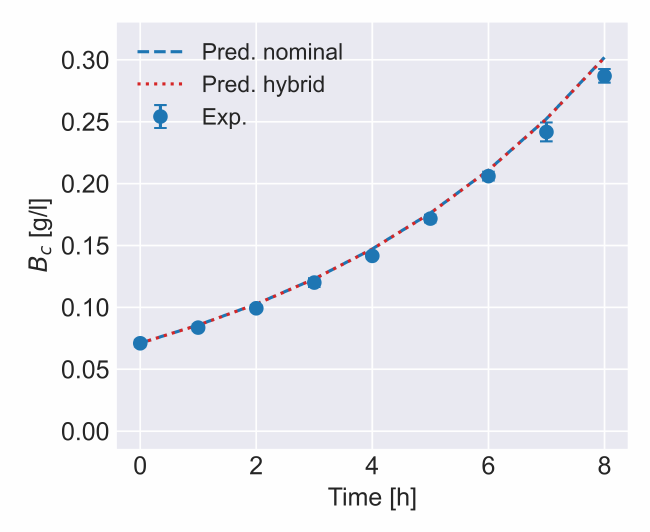}}
\subfloat{\includegraphics[width=0.28\textwidth, keepaspectratio]{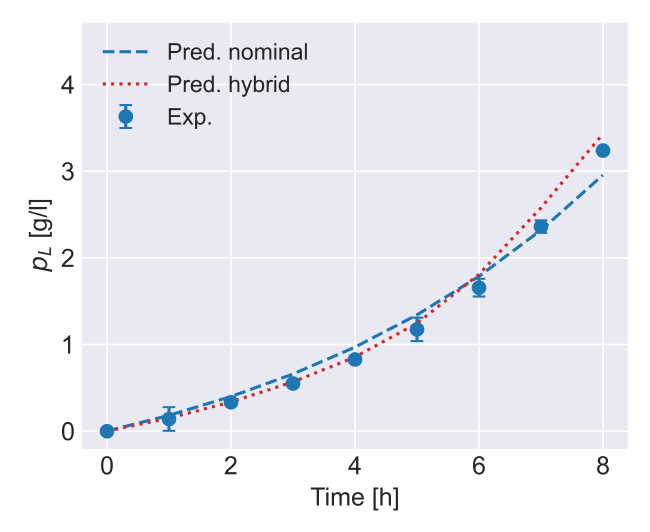}} 
\vspace{-0.30cm}
\begin{flushleft}
\footnotesize{ B) Batch with constant $u_l = 175 \, \mathrm{\mu mol \, m^{-2} \, s^{-1}}$}   
 \end{flushleft}
 \vspace{-0.30cm}
\subfloat{\includegraphics[width=0.28\textwidth, keepaspectratio]{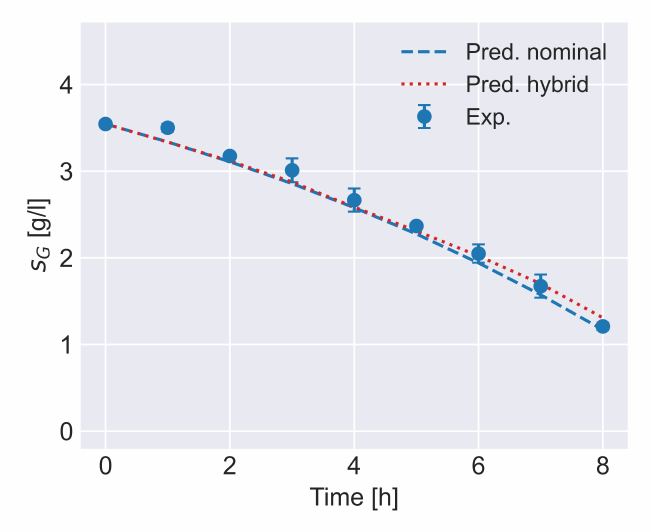}}
\subfloat{\includegraphics[width=0.28\textwidth, keepaspectratio]{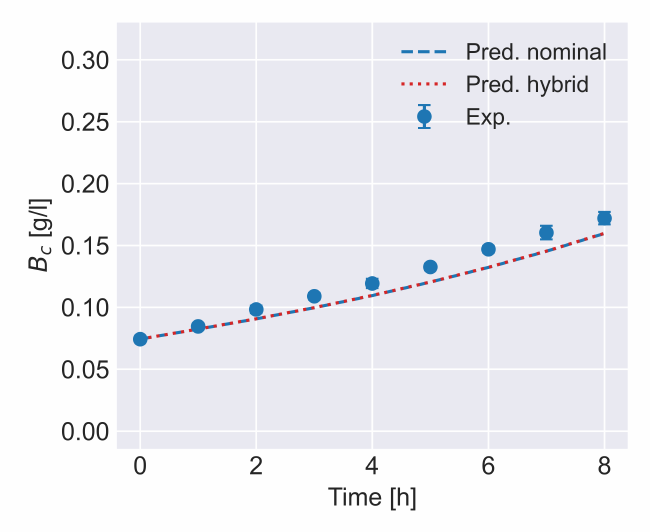}}
\subfloat{\includegraphics[width=0.28\textwidth, keepaspectratio]{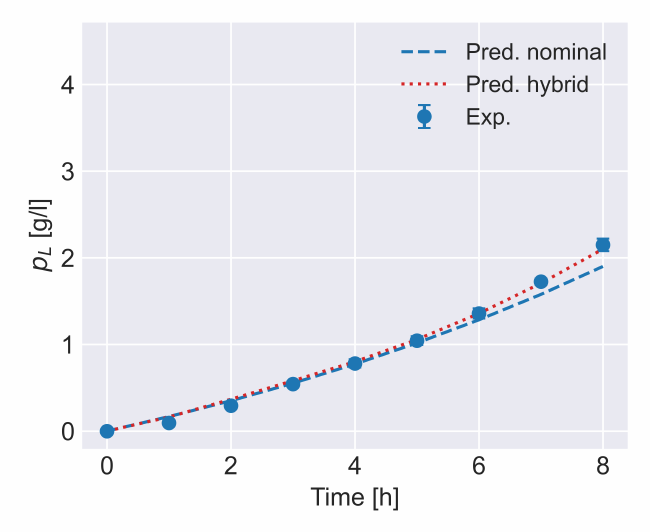}}
\vspace{-0.30cm}
\begin{flushleft}
\footnotesize{C) Batch with constant $u_l = 349 \, \mathrm{\mu mol \, m^{-2} \, s^{-1}}$}
\end{flushleft}
\vspace{-0.30cm}
\subfloat{\includegraphics[width=0.28\textwidth, keepaspectratio]{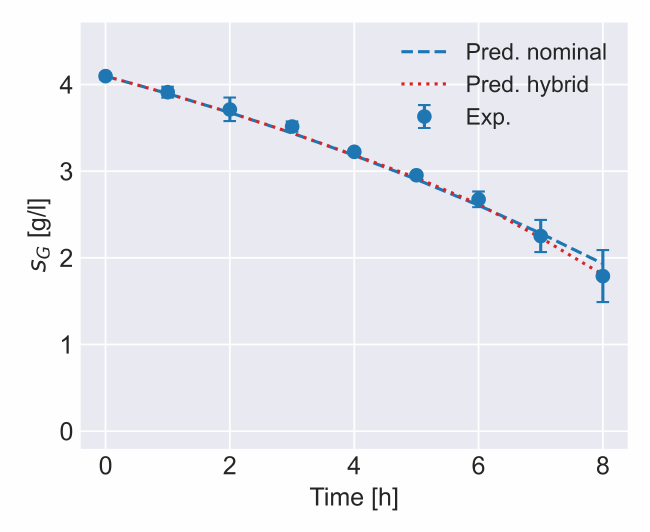}}
\subfloat{\includegraphics[width=0.28\textwidth, keepaspectratio]{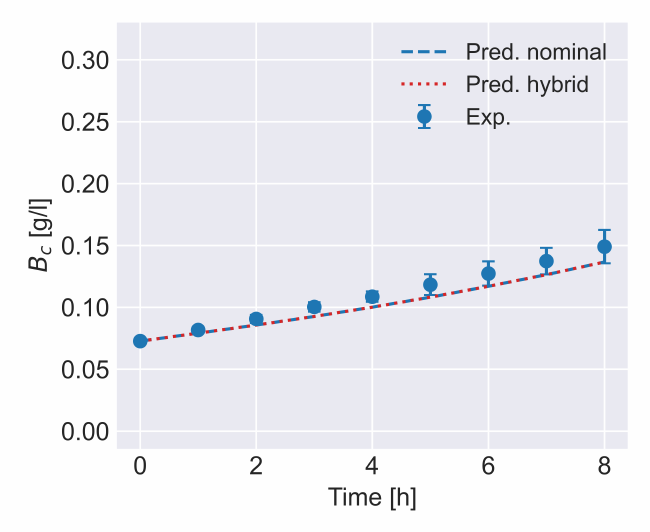}}
\subfloat{\includegraphics[width=0.28\textwidth, keepaspectratio]{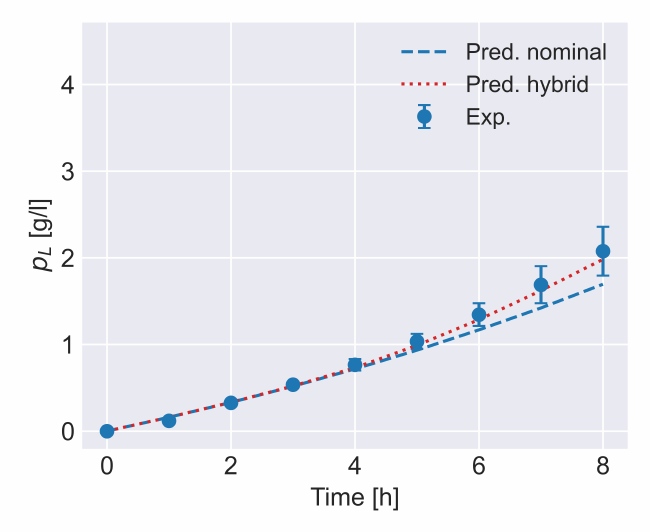}}
\vspace{-0.30cm}
\begin{flushleft} 
\footnotesize{D) Batch with constant $u_l = 524 \, \mathrm{\mu mol \, m^{-2} \, s^{-1}}$}
\end{flushleft}
\vspace{-0.30cm}
\subfloat{\includegraphics[width=0.28\textwidth, keepaspectratio]{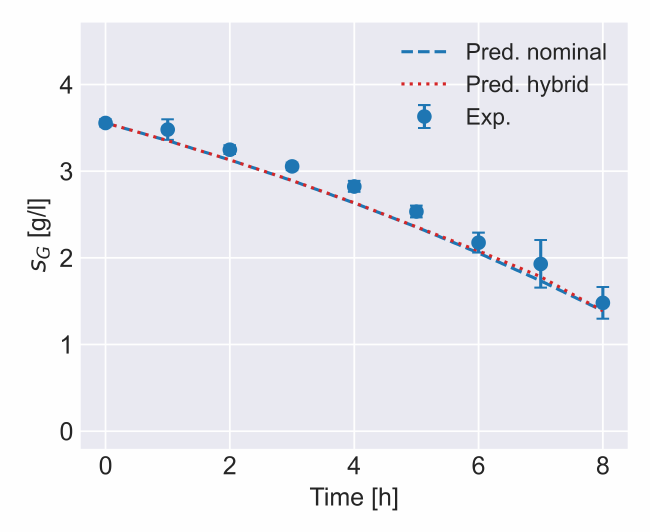}}
\subfloat{\includegraphics[width=0.28\textwidth, keepaspectratio]{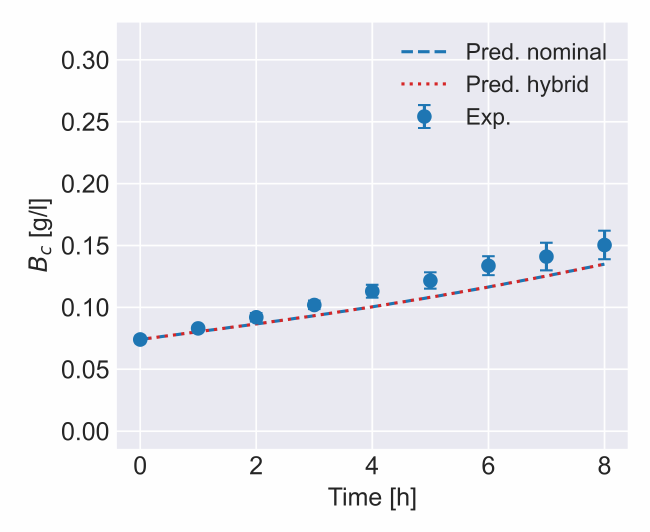}}
\subfloat{\includegraphics[width=0.28\textwidth, keepaspectratio]{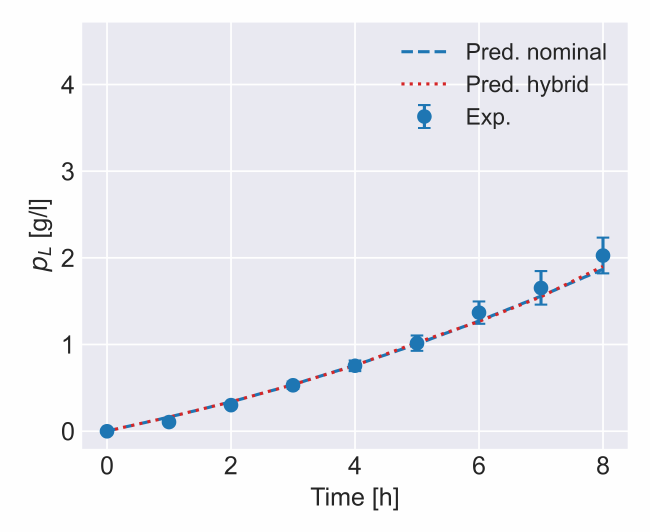}}
\vspace{-0.30cm}
\begin{flushleft} 
\small{E) Batch with constant $u_l = 873 \, \mathrm{\mu mol \, m^{-2} \, s^{-1}}$} 
\end{flushleft}
\vspace{-0.30cm}
\subfloat{\includegraphics[width=0.28\textwidth, keepaspectratio]{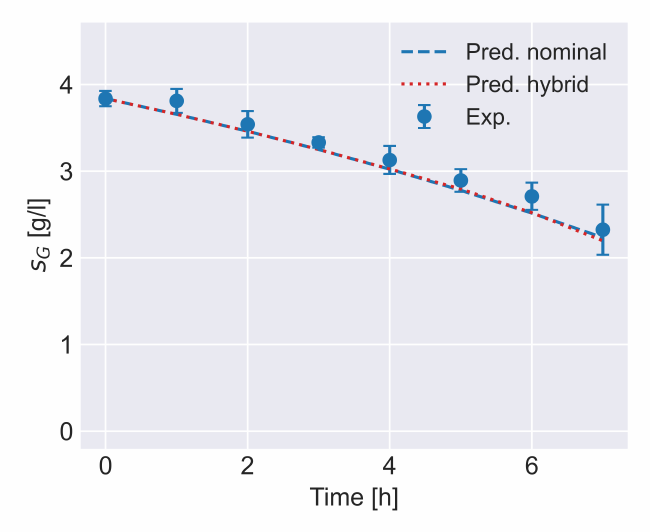}}
\subfloat{\includegraphics[width=0.28\textwidth, keepaspectratio]{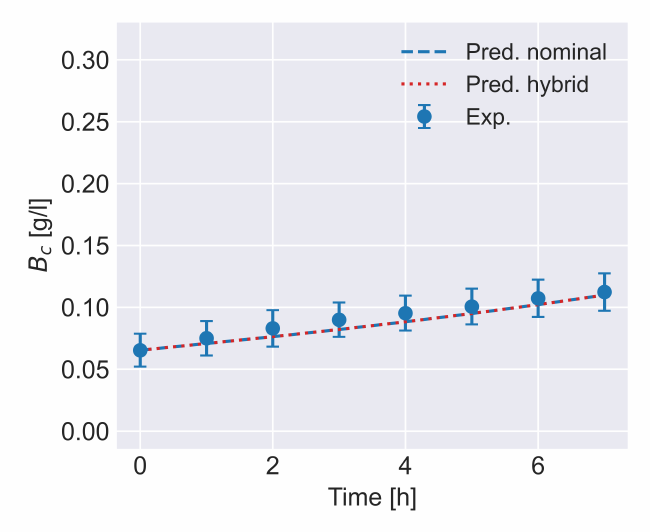}}
\subfloat{\includegraphics[width=0.28\textwidth, keepaspectratio]{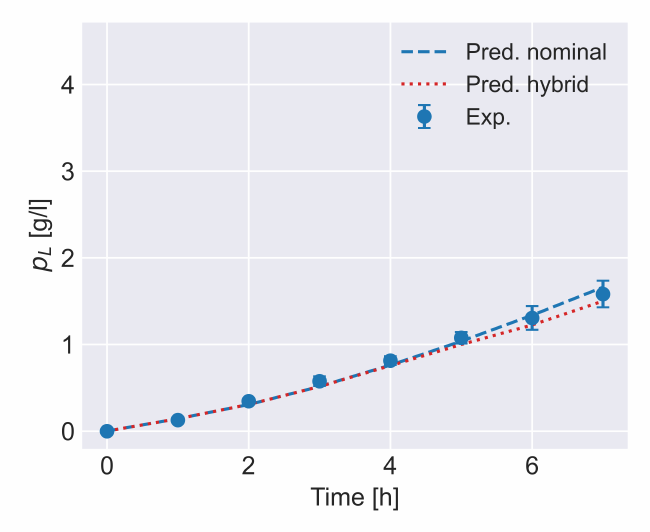}}
\caption{Fitting of the nominal and hybrid models to the experimental data under different constant light inputs.}
\label{fig:estimation}
\end{figure}

Based on the batch experiments presented in Fig. \ref{fig:estimation}, we carried out a parameter estimation procedure considering the data from all the batch experiments simultaneously. The resulting optimized \textit{nominal} parameter values are $k_\mathrm{BV} = 2.605\times 10^{-4} \, \mathrm{VU/g}$, $k_G = 5.340\times 10^{-7} \, \mathrm{g/l}$, $k_\mathrm{GV} = 1.053\times 10^{-6} \, \mathrm{VU/g}$, $k_\mathrm{LV} = 1.002\times 10 \, \mathrm{VU/g}$, $m_G = 1.232\times 10^{-6} \, \mathrm{g/g/h}$, $m_L = 1.910 \, \mathrm{g/g/h}$, $n_1 = 1.000\times 10^{-2}$, $n_2 = 1.028\times 10^{-1}$, $n_3 = 1.000\times 10$, $q_{G_\mathrm{max}} = 1.731 \, \mathrm{g/g/h}$, $Y_\mathrm{BG} = 1.083\times 10^{-1} \, \mathrm{g/g}$, $Y_\mathrm{LB} =  2.204 \, \mathrm{g/g}$, $q_{E_0} = 1.000\times 10^{-6} \, \mathrm{VU/g/h}$, $q_{E_\mathrm{max}} = 1.000\times 10 \, \mathrm{VU/g/h}$, $n_4 = 4.718$, $k_u =  3.729\times 10^{2} \, \mathrm{\mu mol \, m^{-2} \, s^{-1}}$, $k_d =  0.988 \, \mathrm{1/h}$. Note that the latter parameters were estimated considering \textit{only} the knowledge-based part of the model, i.e., without learning the model-plant mismatch with Gaussian processes. We call this model the \textit{nominal} model. Our goal with the parameter estimation was not to provide \textit{unique} parameters rendering \textit{perfect} fitting, but rather to approximate the behavior of the dynamic system such that, if necessary, the model fitting and predictability could be enhanced with Gaussian processes. 

Overall, the nominal model fits well the experimental data for all the tested light inputs in Fig. \ref{fig:estimation}. Despite the good fitting of the nominal model, we still implemented -as a proof of concept- Gaussian process regressors to learn the remaining model-plant mismatch. That is, we learned the model error $w_i$ of equations \eqref{eq:ode_glc}-\eqref{eq:ode_lac}. As inputs for each of the Gaussian process regressors, we used the dynamic input and all the model states, while the output of each of the Gaussian process regressors was the corresponding model-plant error. The model error $w_i$ was computed as
\begin{equation}
w_i(t_k) = \frac{x_e(t_{k+1})-x_e(t_{k})}{t_{k+1}-t_{k}} - \frac{x_m(t_{k+1})-x_m(t_{k})}{t_{k+1}-t_{k}},
\end{equation} 
where $x_e(t_k)$ and $x_e(t_{k+1})$ are the \textit{measured} experimental state values at sampling times $t_k$ and $t_{k+1}$, respectively. Similarly, $x_m(t_k)$ and $x_m(t_{k+1})$ are the \textit{predicted} state values at times $t_k$ and $t_{k+1}$ for the knowledge-based part of the model. Note that at each sampling time, $x_m(t_k) := x_e(t_{k})$. We refer to the model combining the knowledge-based part with the Gaussian process regressors as the \textit{hybrid} model. As expected, the hybrid model can fit slightly better the experimental data, particularly for the glucose and lactate profiles over the mid-end term of the fermentation experiments (cf. Fig. \ref{fig:estimation}).

\subsection{Implementation of open-loop optogenetic control of ATPase}
\label{subsec:opti_results}
We formulated optimal control problems, as described in Section \ref{sec:opt_control}, constrained by the nominal and hybrid models. The idea was to show the applicability of both models for optimal control of the optogenetically controlled ATPase expression. Specifically, we maximized the final batch lactate concentration in a time frame of eight hours, by applying piece-wise constant inputs of green light every hour (\textit{dynamic} degree of freedom) and determining the optimal initial glucose concentration (\textit{static} degree of freedom). Discretizing the input renders the optimization problem finite-dimensional and practical to solve since finding $u_l$ as a \textit{function} would otherwise make the problem infinite-dimensional. The input was constrained to the values used for fitting/training the model (cf. Eq. \eqref{eq:u_cons_cs}) and we demanded the optimizer to deplete all glucose by the end of the fermentation (cf. Eq. \eqref{eq:s_G_cons_cs}). Furthermore, we constrained the optimizer to achieve a user-defined product on glucose yield over the batch (cf. Eq. \eqref{eq:Ylg_cons_cs}) $\Tilde{Y}_\mathrm{LG,batch}$. We also included a box constraint for the initial glucose concentration (cf. Eq. \eqref{eq:box_sg_0}) with zero and $s_{G_\mathrm{max}}=5$ g/l as the lower and upper bounds, respectively.

The resulting optimization problem reads
\begin{maxi!} 
    {u_l(\cdot),s_G(t_0)}{p_L(t_f),\label{eq:optimal_cost_}}{\label{eq:optimal_cs}}{}
    \addConstraint{}{\text{Eqs.}\eqref{eq:ode_glc}-\eqref{eq:ode_E},}{}
    \addConstraint{}{0 \, \leq u_l \leq 873, \label{eq:u_cons_cs}}{}
    \addConstraint{}{s_G(t_f)=0, \label{eq:s_G_cons_cs}}{}
    \addConstraint{}{\frac{p_L(t_f)-p_L(t_0)}{s_G(t_0)-s_G(t_f)} = \Tilde{Y}_\mathrm{LG,batch},\label{eq:Ylg_cons_cs}}{}
    \addConstraint{}{ 0 < s_G(t_0) \leq s_{G_\mathrm{max}}. \label{eq:box_sg_0}}{}
\end{maxi!}

The optimization problem in \eqref{eq:optimal_cs} thus maximizes the volumetric productivity in the fixed time frame under given product yield, light, initial substrate, and substrate consumption constraints. Remark that we neglect the model error $\bm{w}$ from the Eqs. \eqref{eq:ode_glc}-\eqref{eq:ode_E} when we use the \textit{nominal} model in the optimization. When we use instead the \textit{hybrid} model, we then include the Gaussian-process-based model error $\bm{w}$ in Eqs. \eqref{eq:ode_glc}-\eqref{eq:ode_E}. For the optimization based on the nominal model, we arbitrarily set $\Tilde{Y}_\mathrm{LG,batch}$ to 0.954 g/g, while this was set to 0.986 g/g for the optimization based on the hybrid model. The goal was not to compare one-to-one the performance of the optimization using the two models but rather to show the flexibility of our approach to handle both knowledge-based and hybrid Gaussian-supported models. Furthermore, we also wanted to highlight the fact that modulating ATPase expression dynamically can be exploited to adjust the batch-to-batch fermentation performance in terms of product yield and productivity. Hence, the different selected values for $\Tilde{Y}_\mathrm{LG,batch}$.

\begin{figure}[H]
\centering
\begin{center}
\vspace{-0.50cm}
\footnotesize{A) Light input}
\vspace{-0.30cm}
\end{center}
\subfloat{\includegraphics[width=0.3\textwidth, keepaspectratio]{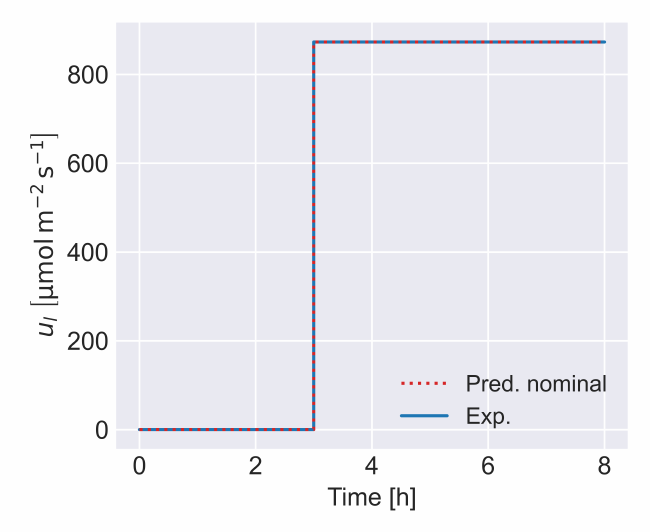}}
\subfloat{\includegraphics[width=0.3\textwidth, keepaspectratio]{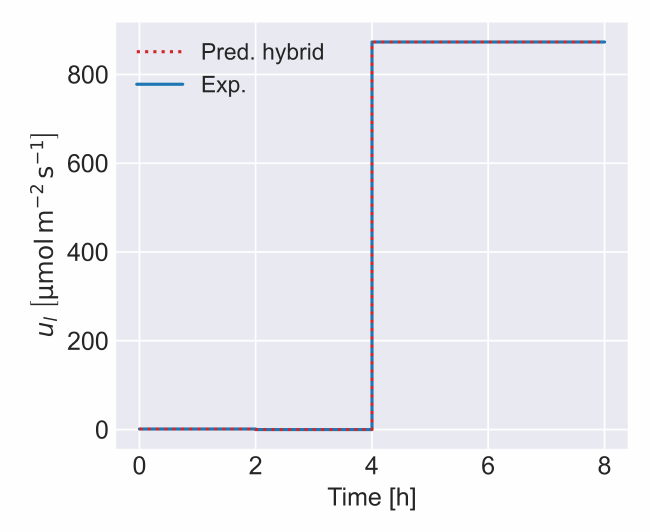}}
\centering
\begin{center} 
\vspace{-0.50cm}
\footnotesize{B) Glucose}
\vspace{-0.30cm}
\end{center}
\subfloat{\includegraphics[width=0.3\textwidth, keepaspectratio]{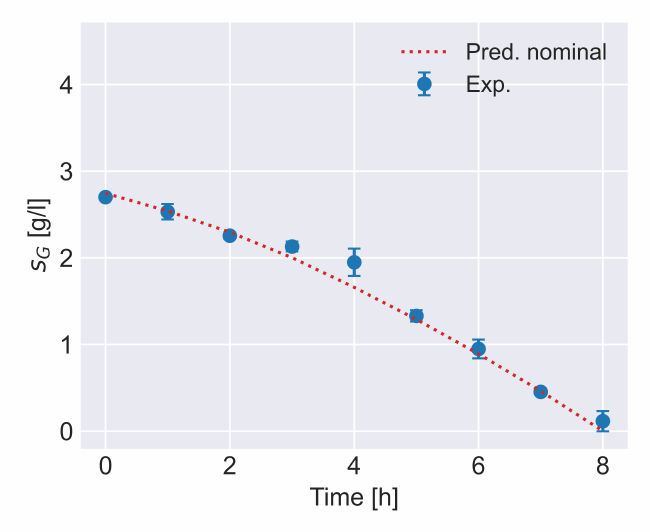}} 
\subfloat{\includegraphics[width=0.3\textwidth, keepaspectratio]{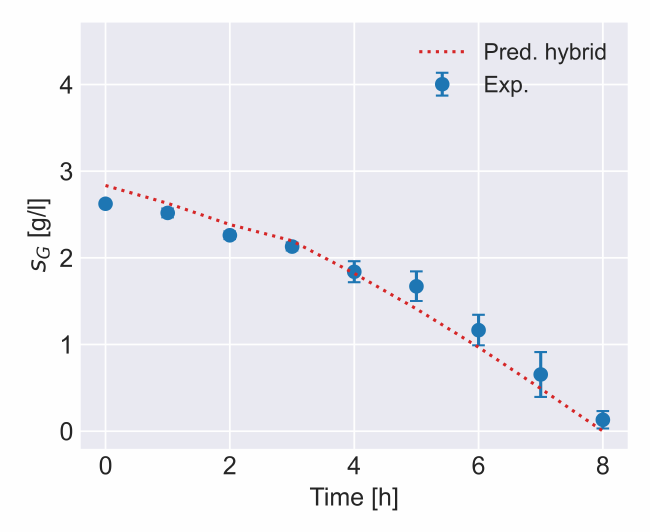}}
\begin{center} 
\vspace{-0.50cm}
\footnotesize{C) Biomass}
\vspace{-0.30cm}
\end{center}
\subfloat{\includegraphics[width=0.3\textwidth, keepaspectratio]{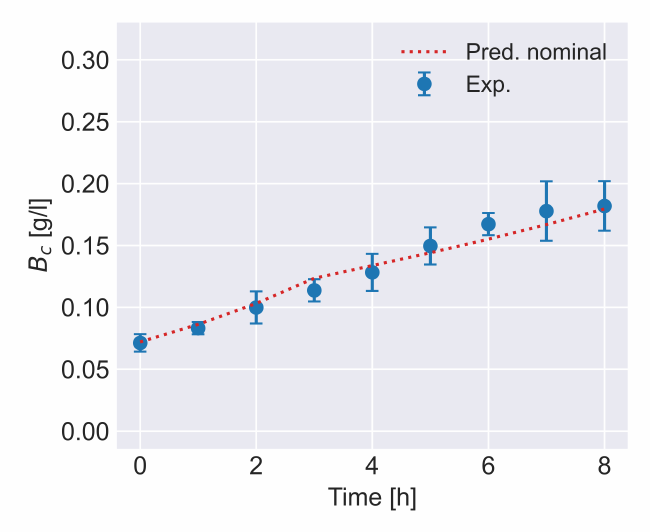}} 
\subfloat{\includegraphics[width=0.3\textwidth, keepaspectratio]{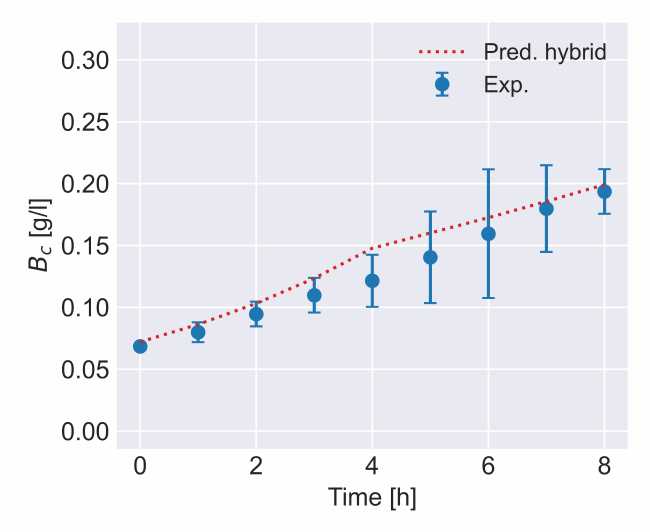}}
\begin{center} 
\vspace{-0.50cm}
\footnotesize{D) Lactate}
\vspace{-0.30cm}
\end{center}
\subfloat{\includegraphics[width=0.3\textwidth, keepaspectratio]{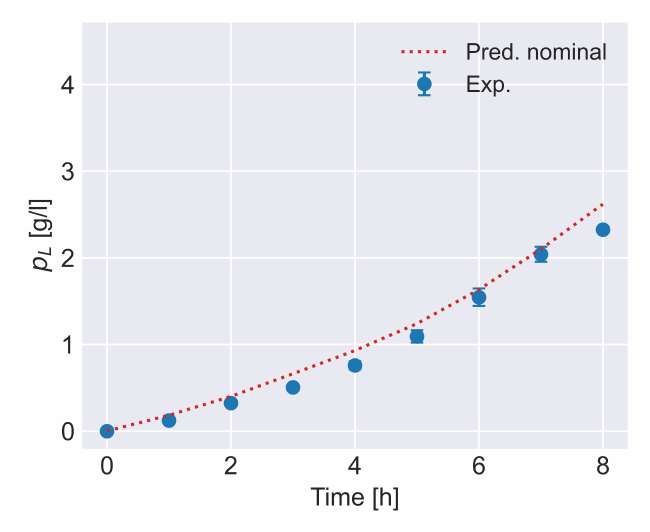}}
\subfloat{\includegraphics[width=0.3\textwidth, keepaspectratio]{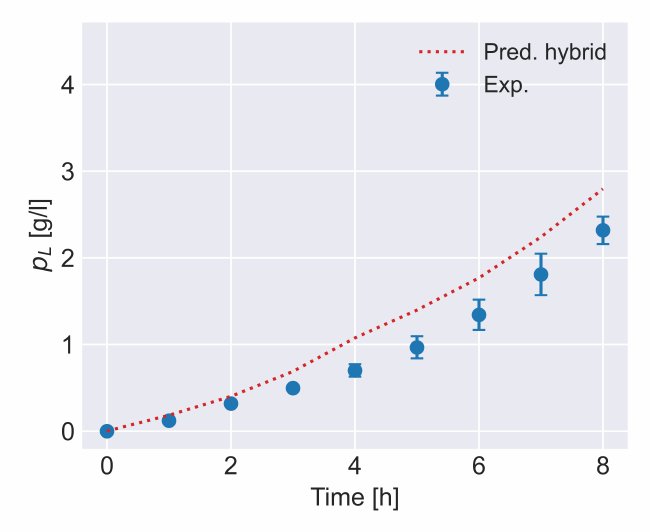}}
\begin{center} 
\vspace{-0.50cm}
\footnotesize{E) (Virtual) ATPase}
\vspace{-0.30cm}
\end{center}
\subfloat{\includegraphics[width=0.3\textwidth, keepaspectratio]{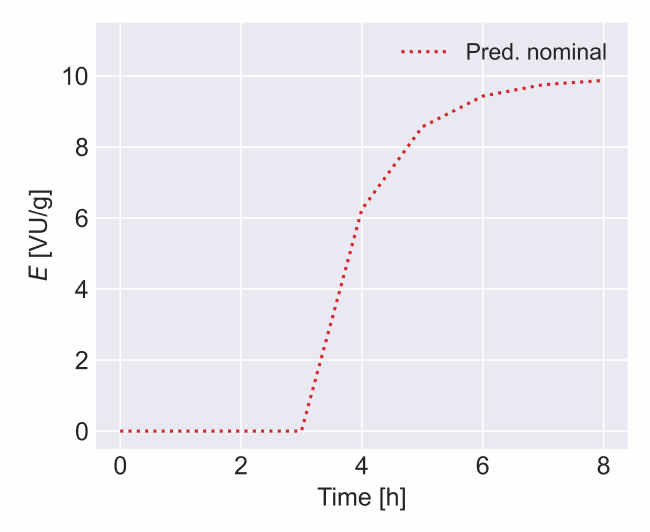}}
\subfloat{\includegraphics[width=0.3\textwidth, keepaspectratio]{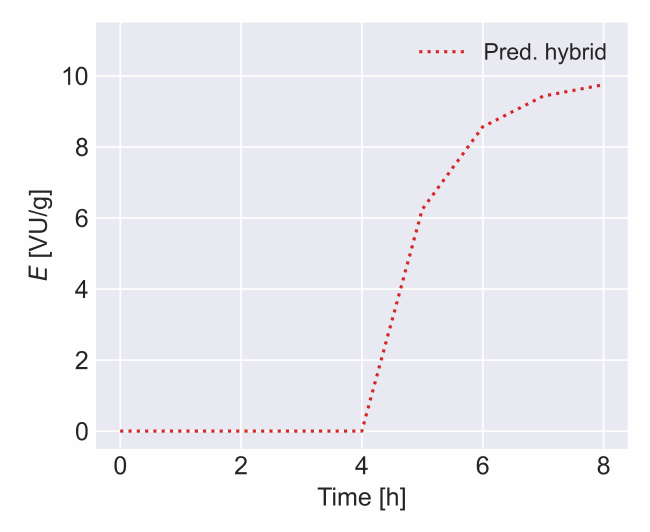}}
\caption{Results of the open-loop optimization prediction and experimental implementation. The optimizations based on the nominal and hybrid models are shown on the left and right sides, respectively.}
\label{fig:olo_pred_exp}
\end{figure}

Intuitively, as long as $\Tilde{Y}_\mathrm{LG,batch}$ is larger than the one achievable by the cell without ATPase induction (cf. Fig. \ref{fig:fermentation_metrics}, constant $u_l=0$), the optimizer is expected to utilize the light-mediated ATP turnover mechanism to increase the product yield. Therefore, formulating an optimal control problem as done in \eqref{eq:optimal_cs} is a way to obtain trade-offs between enhancement of product yield and drop in volumetric productivity in the context of dynamic ATP turnover as discussed in \cite{espinel_ATP_w_2022,espinel_opt_2022}. The predicted open-loop optimization results based on the nominal and hybrid models are presented in Fig. \ref{fig:olo_pred_exp}. 

In both open-loop optimizations, the predicted input follows a two-stage profile, with a first phase at 0 $\mathrm{\mu mol \, m^{-2} \, s^{-1}}$ (no ATPase induction), followed by a second phase at 873 $\mathrm{\mu mol \, m^{-2} \, s^{-1}}$ (induction at the maximum green light photon flux density value). The main difference between the two optimization problems is the time at which the second phase is triggered, i.e., 3 h for the optimization based on the nominal model and 4 h for the optimization based on the hybrid model. As expected, in both cases the optimizer predicts a slight decrease in the biomass growth rate and a slight increase in the glucose uptake rate and lactate production rate for the second fermentation phase. The predicted optimal initial glucose concentration determined by the optimization based on the nominal model is 2.745 g/l, while this is 2.834 g/l for the optimization based on the hybrid model. As demanded, the optimizer predicts full consumption of glucose by the end of the fermentations.

The predicted yields of lactate on glucose over the batch for both fermentations are as demanded, i.e., 0.954 g/g for the optimization based on the nominal model and 0.986 g/g for the optimization based on the hybrid model (Fig. \ref{fig:fermentation_metrics}, orange bars). \textit{Compared to the scenario without ATPase induction} (Fig. \ref{fig:fermentation_metrics}, $u_l=0$), this represents a predicted increase in product yield of 9 \% and 13 \%, respectively. The predicted increase in batch product yield is at the expense of a decrease in the biomass on glucose yield by 33 \% and 23 \%, respectively. This also correlates to a decrease in volumetric productivity by 19 \% and 14 \%, respectively. As can be seen, the hybrid model predicts a less pronounced drop in biomass on glucose yield and volumetric productivity, even though the demanded increase in product yield in the optimization with the hybrid model is higher than in the optimization with the nominal model. In other words, the hybrid model seems to be more \textit{optimistic}, which is in line with the fittings observed in Fig. \ref{fig:estimation}, where the hybrid model predicts slightly higher lactate formation rates (in particular for treatments at 0, 175, 349 $\mu$mol$\ $m$^{-2}\ $s$^{-1}$). 

We experimentally validated the predicted open-loop optimizations. As can be seen in Fig. \ref{fig:olo_pred_exp}, the experiments follow the overall trends as predicted by the optimal control problems. There is, nevertheless, some model-plant mismatch, in particular for the lactate concentration profile in the optimization based on the hybrid model. One should also notice that there is a slight mismatch between the target and the experimental initial concentrations, which is natural due to human error (cf. e.g. the glucose initial concentration in Fig. \ref{fig:olo_pred_exp}, optimization based on the hybrid model). The final lactate concentrations for both optimizations are within the same range, i.e., $2.324 \pm 0.020$ and $2.317	\pm 0.018 \, \mathrm{g/l}$ for the optimizations based on the nominal and hybrid models, respectively. \textit{Compared to the scenario without ATPase induction} ($u_l=0$ constant), the increase in lactate on glucose yield over the entire batch for the optimizations based on the nominal and hybrid models is, in \textit{average}, 3 \% and 6 \%, respectively (cf. Fig. \ref{fig:fermentation_metrics}, green bars). In both optimizations, this is lower than the demanded $\Tilde{Y}_\mathrm{LG,batch}$ values; however, if we take into account the reported standard deviations, the experimental results are still \textit{close} to the predicted values. 

It is worth noting that the predicted input, following an off-on trajectory, differs from the more \textit{gradual} trajectories found in previous simulation studies involving bilevel optimizations using dynamic constraint-based models (cf. \cite{espinel_opt_2022,espinel_cyb_fram_2023}). This difference can be explained by the fact that the dynamic constraint-based models mentioned earlier are constrained by redox and energy balances, as well as resource allocation phenomena, which are neglected in our simplified modeling approach. Also, constraint-based models as in \cite{espinel_opt_2022,espinel_cyb_fram_2023} assume a cell's dynamic (\textit{evolutionary}) objective function, which, when formulated within a (bilevel) optimization scheme may bias the outcome of the optimization. Overall, we deem the predicted inputs in this work reasonable since they follow the widely discussed two-stage fermentation approaches for bioprocess optimization \cite{klamt_when_2018,burg_large-scale_2016,lalwani_current_2018}. Future experimental comparison of optimal control using a \textit{validated} constraint-based dynamic model and a macro-kinetic-like model as the one outlined here is, nevertheless, of interest to elucidate whether a higher modeling complexity also translates into enhanced predictability for model-based optimization and control.

The fact that the optimizations were performed in an open-loop fashion, i.e., without taking correcting actions online, hinders the full potential of model-based optimization. To address possible system uncertainty, as observed in the validation experiments, one could, e.g., implement feedback control schemes such as model predictive control. In model predictive control, the optimization is updated with the initial conditions of the plant and resolved at every sampling point (cf. e.g. \cite{jabarivelisdeh_optimization_2018,morabito_multi-mode_2019,jabarivelisdeh_adaptive_2020,morabito_towards_2021,morabito_efficient_2022,espinel_opt_2022,espinel_cyb_fram_2023,espinel-rios_hybrid_2024}). The repetitive solution of the optimization problem leads to closed-loop control. Closed-loop control when considered in the frame of metabolic systems with external induction of gene expression is often referred to as \textit{metabolic cybergenetics} \cite{carrasco-lopez_optogenetics_2020,espinel_cyb_fram_2023,espinel-rios_hybrid_2024}. 

Other modeling techniques for metabolic systems with external induction of gene expression are also worth considering for future experimental implementations of model-based control. One interesting approach is the use of flux balance analysis to \textit{inform} machine-learning surrogates to be embedded into the reaction rates of macro-kinetic models, effectively creating hybrid \textit{physics-informed} models \cite{espinel-rios_linking_2023,espinel-rios_hybrid_2024}. In this case, valuable information captured by metabolic networks, such as intrinsic metabolic trade-offs, redox and energy balances, etc., can be transferred into structurally simpler models. This strategy could help to reduce the gap between unstructured kinetic models and structured models based on flux balance analysis. The use of Gaussian processes to learn the model error, as described in this study, could still be employed to enhance the predictability of these hybrid physics-informed models. Finally, even though in this work we only considered the mean of Gaussian processes, one could in principle also leverage the predicted variance of the Gaussian processes to balance \textit{exploration} and \textit{exploitation} properties in the optimization problems. That is, one could use the predicted variance(s) in the objective function of the optimization to actively explore areas with high uncertainty while still exploiting the system \cite{espinel-rios_batch--batch_2023}.

\section{Conclusion}
\label{sec:conclusion}
We have proposed and \textit{experimentally validated} a (Gaussian-process-supported) model-based optimization strategy for open-loop optogenetic control of the ATPase to maximize bioprocesses production efficiency through dynamic enforced ATP turnover. We have outlined a simplified modeling framework, i.e., an unsegregated and quasi-unstructured kinetic modeling approach, that captures relevant process dynamics. This facilitates model parameterization and simplifies model-based optimization compared to previously proposed dynamic constraint-based models. We have also considered hybrid models combining knowledge-based and Gaussian-process-supported components for modeling and optimal control.

For the experimental implementation, we have engineered \textit{E. coli} to carry the CcaS/CcaR system to achieve optogenetic control of the ATPase. This engineered \textit{E. coli} produces lactate as the main fermentation product under anaerobic conditions. Following optimal control problems constrained by knowledge-based and hybrid models, we have maximized lactate concentration while aiming at a target product yield and depleting all available glucose. However, there is still some model-plant mismatch which limits the full potential of the presented approach. Further work includes the experimental implementation of model predictive control schemes coupled with soft sensors in the context of metabolic cybergenetic systems. Overall, the presented model-based open-loop optimization strategy, validated with experiments, outlines an example of a simple and structured way to maximize production efficiency in optogenetically regulated metabolic processes. 

\section{Author contributions}
\textbf{Sebastián Espinel-Ríos}: Conceptualization, Methodology, Software, Formal Analysis, Writing - Original Draft, Writing - Review \& Editing, Visualization.
\textbf{Gerrich Behrendt}: Methodology, Investigation, Writing - Original Draft, Writing - Review \& Editing, Visualization.
\textbf{Jasmin Bauer}: Investigation, Validation, Writing - Original Draft, Writing - Review \& Editing.
\textbf{Bruno Morabito}: Methodology, Writing - Original Draft, Writing - Review \& Editing.
\textbf{Johannes Pohlodek}: Methodology, Writing - Original Draft, Writing - Review \& Editing.
\textbf{Andrea Schütze}: Investigation, Validation, Writing - Review \& Editing.
\textbf{Rolf Findeisen}: Conceptualization, Supervision, Writing - Review \& Editing.
\textbf{Katja Bettenbrock}: Conceptualization, Writing - Original Draft, Writing - Review \& Editing, Writing - Review \& Editing, Supervision, Funding Acquisition.
\textbf{Steffen Klamt}: Conceptualization, Writing - Original Draft, Writing - Review \& Editing, Supervision, Funding Acquisition.

\section{Acknowledgment}
This work was supported by the International Max Planck Research School for Advanced Methods in Process and Systems Engineering (IMPRS ProEng). We would like to thank Reiner Könning for his support with the instrumentation of the light-delivery system.\newline

\noindent \textbf{Declarations of interest} \newline
The authors declare that they have no conflict of interest. \newline

\noindent \textbf{Data accessibility} \newline
The data that support the findings of this study are available from the corresponding author upon reasonable request.

\bibliographystyle{elsarticle-num} 
\bibliography{bibliography}

\end{document}